\documentclass[lettersize,journal]{IEEEtran}
\usepackage{amsthm}
\usepackage{amsmath,amsfonts}
\usepackage{algorithmic}
\usepackage{algorithm}
\usepackage{amssymb}
\usepackage{array}
\usepackage[caption=false,font=normalsize,labelfont=sf,textfont=sf]{subfig}
\usepackage{textcomp}
\usepackage{stfloats}
\usepackage{url}
\usepackage{verbatim}
\usepackage{graphicx}
\usepackage{cite}
\usepackage{cases}
\usepackage{hyperref}
\usepackage{bm}
\usepackage{color}
\usepackage{gensymb}
\usepackage{multirow}
\usepackage{booktabs}
\usepackage{makecell}
\usepackage{threeparttable}
\usepackage{utfsym}
\usepackage[most]{tcolorbox}
\DeclareMathOperator*{\argmax}{arg\,max}

\hypersetup{colorlinks=true,
	linkcolor=blue,
	citecolor=blue,      
	urlcolor=black,
}

\newtheoremstyle{mylemma}
{}{}                 
{\normalfont}        
{}                   
{\bfseries}         
{.}                  
{ }                  
{\thmname{#1}\thmnumber{ #2}\thmnote{ (#3)}} 

\theoremstyle{mylemma}

\newtheorem{proposition}{Proposition}

\begin{document}

\title{Strategic Beam Brainstorming for Generative Site-Specific Beamforming}

\author{Zihao Zhou,~\IEEEmembership{Graduate Student Member,~IEEE}, Zhaolin Wang,~\IEEEmembership{Member,~IEEE}, \\ and Yuanwei Liu,~\IEEEmembership{Fellow,~IEEE}
\thanks{The authors are with the Department of Electrical and Computer Engineering, The University of Hong Kong, Hong Kong (e-mail: eezihaozhou@connect.hku.hk,zhaolin.wang@hku.hk,yuanwei@hku.hk)}}

\maketitle

\begin{abstract}
A novel plug-and-play candidate beam selection strategy is proposed for generative site-specific beamforming (GenSSBF), a paradigm that draws on a learned conditional channel distribution to generate multiple plausible beams. Under a limited probing budget, realizing the full potential of GenSSBF requires \textit{identifying which candidate beams are worth probing}. To this end, the marginal probing value is first derived to guide strategic candidate probing by quantifying each beam’s additional gain beyond those already selected. Based on this criterion, two algorithms are developed without retraining the generator: 1) Greedy candidate probing (GCP) selects all candidate beams from a locally generated pool prior to over-the-air probing, requiring only a single feedback round; 2) Adaptive candidate probing (ACP) exploits reference signal received power (RSRP) feedback after each probe to update channel hypothesis weights and guide the next beam selection, without increasing the asymptotic complexity. Simulations across three ray-tracing scenarios under noiseless and noisy probing reveal that: 1) candidate complementarity is key to efficient beam brainstorming; and 2) ACP is particularly beneficial when initial sensing leaves substantial channel uncertainty and subsequent feedback is reliable.
\end{abstract}

\vspace{-1.2em}
\section{Introduction}
\IEEEPARstart{A}{chieving} high beamforming gain with limited channel probing is a central challenge in wireless systems with large antenna arrays \cite{xiao2017millimeter}. While exhaustive beam sweeping identifies the best beam in a predefined codebook, its measurement overhead scales with the codebook size \cite{tan20024beam}. Site-specific beamforming (SSBF) reduces this overhead by exploiting recurring propagation patterns at a deployment site to infer promising beams from only a few initial probes\cite{heng2024site}.

However, coarse initial reference signal received power (RSRP) measurements are often ambiguous, as distinct channels can produce similar observations while favoring substantially different beams. Consequently, a conventional point prediction-based deep learning (DL) model may average across these plausible beamforming vectors \cite{torralba2024foundations}, yielding a beam that is poorly matched to the actual channel. Recently, generative site-specific beamforming (GenSSBF) has been proposed to address this \textit{ambiguity} by learning the conditional distribution of the channel given the initial measurements \cite{wang2026generative}. A distinctive feature of GenSSBF is its ability to ``\textit{brainstorm}''. Given the initial measurements, it draws on learned site-specific knowledge to propose several plausible beams. A limited number of additional probes are then used to determine the beam that performs best on the actual communication link.

\subsection{Prior Work}
\subsubsection{Site-Agnostic Beamforming}
Conventional codebook-based beamforming typically follows a ``probing-measuring-reporting'' procedure using a predefined codebook. Specifically, the base station (BS) transmits reference signals using selected codebook beams, while the user equipment (UE) measures the corresponding received signal quality and provides feedback for beam selection. The effectiveness of this procedure thus depends on both the codebook itself and the probing overhead involved in identifying a suitable beam. Discrete Fourier transform (DFT) codebooks \cite{love2003equal} provide a simple way to construct constant-modulus steering beams to achieve angular coverage. Oversampled DFT (O-DFT) codebooks offer a denser grid of beam directions to reduce the spatial quantization loss, at the expense of increased beam sweeping overhead. To reduce this overhead, a hierarchical codebook organized as a binary tree was developed in \cite{he2015suboptimal}, enabling coarse-to-fine beam training. By exploiting subarray processing and antenna deactivation, a new method to construct hierarchical beams with different widths was proposed \cite{xiao2016hierarchical}. Noting that channel sounding and data transmission impose conflicting requirements on beam design, i.e., wide beams reduce sounding overhead, whereas narrow beams provide higher beamforming gain, the authors of \cite{song2017common} designed a common codebook for uniform planar arrays (UPAs), approximating ideal flat-top beam patterns over predefined angular regions. Focusing on random device orientations, a codebook refinement method was proposed in \cite{bozkurt2025beam}, aiming to reduce the codebook size while keeping the beamforming gain loss within a prescribed threshold at any orientations of the user terminal. In practice, at a given deployment site, the surrounding geometry gives rise to recurring propagation patterns, making some beam directions more likely to provide strong received signals than others. However, the aforementioned methods generally do not explicitly exploit prior knowledge of these patterns when designing codebooks or search strategies. Consequently, this motivates SSBF, which incorporates site-specific propagation knowledge to achieve a more favorable tradeoff between beamforming performance and channel probing overhead.

\subsubsection{Site-Specific Beamforming}
SSBF exploits the propagation characteristics of a particular deployment site to reduce the measurements required for beamforming. In \cite{alrabeiah2022neural}, a complex-valued neural network (NN) was used to represent a probing codebook, with its trainable weights directly determining the beamforming vectors. Joint learning of the probing codebook and beam selector was subsequently investigated through an end-to-end framework in \cite{heng2022learning}, with a hierarchical beam alignment extension developed in \cite{yang2024hierarchical}. Reinforcement learning (RL) provides another approach to codebook adaptation: the methods in \cite{zhang2022reinforcement}, \cite{abdallah2024multiagent} update beam patterns through online interactions with the environment using received signal power as feedback. Despite adapting the codebook to the deployment site, restricting the final beam to a finite set of candidates can still incur grid mismatch. To overcome this limitation, the authors of \cite{heng2024grid} developed a grid-free framework that jointly learns a probing codebook and a multilayer perceptron (MLP) for direct beam synthesis. A fundamental challenge in SSBF is the \textit{ambiguity} inherent in coarse channel observations: similar measurements may correspond to distinct channel realizations with substantially different optimal beams. Point prediction does not explicitly retain these alternative solutions. For instance, beam regression trained with a mean-squared-error (MSE) loss favors the conditional mean, potentially producing a compromise that poorly matches the plausible channel realizations \cite{wang2026generative}. More generally, committing to a single beam under ambiguous observations risks a mismatch with the actual channel. 

\subsubsection{Generative Site-Specific Beamforming}
GenSSBF addresses this uncertainty by learning a conditional distribution over plausible beamforming solutions given the coarse observations. Candidate beams are sampled from the learned distribution and evaluated through additional probing to identify a suitable beam for the actual channel. An early realization is beam-brainstorm (BBS) \cite{zhou2026beam}, which employs a conditional diffusion model to learn the channel distribution given the RSRP measurements from a small number of DFT probes. The model generates multiple plausible beams, which are then evaluated through additional probing. Building on this idea, fast BBS (F-BBS) was developed in \cite{zhou2026fastbeambrainstorm} to enable one-step beam generation by learning an average velocity field for the generative process. A masking mechanism was further introduced to accommodate variable-length probing observations. The GenSSBF paradigm was then expanded to UPA systems in \cite{tang2026generative}, where azimuth and elevation probing were decoupled to reduce sensing overhead. A cross-attention encoder was designed to fuse these separate RSRP observations, conditioning a normalized flow model to generate candidate beams for subsequent probing. Beyond standard DFT probing, the authors of \cite{zhao2026generative} investigated a site information-maximizing (SIM) codebook by maximizing the mutual information between RSRP feedback and the site-specific channel. GenSSBF was further extended to multi-user settings in \cite{jiang2026kdgen-bf}, where a conditional diffusion transformer was trained to jointly generate a multi-user beamforming matrix from low-dimensional RSRP observations while accounting for inter-user interference. To support beam prediction under user mobility, Beam-STeP was proposed in \cite{wang2026beamstep}, using a history-conditioned diffusion model to learn the distribution of future beamforming vectors by exploiting site-specific propagation and mobility priors alongside historical beam observations.

\subsection{Motivation and Contributions}
The GenSSBF methods discussed above exploit \textit{beam brainstorming} to translate the learned conditional distribution into improved beamforming performance. Given coarse initial channel observations, a generative model produces multiple candidate beams, which are subsequently probed to identify the one that performs best on the actual communication link. Exploring several plausible beams increases the chance of finding a suitable beam for the actual channel. This benefit, however, comes with additional probing overhead, i.e., each candidate evaluated over the air requires additional probing resources. Therefore, under a limited probing budget, the effectiveness of beam brainstorming depends on which candidates are probed.

A natural implementation is to independently sample candidate beams from the trained generative model and probe all of them. Although each candidate may be plausible given the initial observations, independently generated candidates can offer overlapping gains across possible channel realizations. For example, two candidates may both perform well on the actual channel, once one of these candidates has been included, probing the other may yield little additional benefit. Such redundancy can arise even when the generative model accurately captures the conditional distribution, because independent sampling does not account for the contribution of a candidate relative to the others. Consequently, a key question is \textit{how to determine which candidate beams are worth additional probing under a limited probing budget}, which has received limited attention in existing GenSSBF studies.

The main contributions are summarized as follows:
\begin{itemize}
	\item We formulate beam brainstorming in GenSSBF as a candidate selection optimization problem under a limited probing budget. A marginal probing value is first derived to quantify the additional gain a candidate provides beyond the beams already selected, explicitly accounting for their overlapping gains across plausible channels. Based on this criterion, we develop a plug-and-play greedy candidate probing (GCP) algorithm that selects candidates from a locally generated pool without retraining the generative model. We also analyze the structural properties and the computational complexity of the proposed GCP algorithm.
	
	\item We further propose an adaptive candidate probing (ACP) algorithm that exploits the RSRP feedback obtained during candidate probing to guide subsequent beam selection. Specifically, ACP updates the weights of generated channel hypotheses according to the consistency between their predicted responses and the reported measurements, and selects the next candidate based on its weighted marginal probing value. By alternating between beam selection, online probing, and hypothesis reweighting, ACP incorporates accumulated feedback into the assessment of each candidate's potential gain improvement.
	
	\item We evaluate the proposed methods in three ray-tracing (RT) scenarios under both noiseless and noisy probing. The results demonstrate that: 1) both GCP and ACP improve beamforming gain over independent sampling at the same budget, with ACP capable of achieving higher gain using 50\% fewer candidate probes; 2) the benefit of ACP depends on the uncertainty remaining after initial probing and the reliability of subsequent feedback; 3) larger candidate pools improve beamforming performance at the cost of additional local computation.
\end{itemize}

The remainder of this paper is organized as follows. The system model and problem formulation are described in Section \ref{section2}. Section \ref{section3} develops the GCP algorithm and analyzes its theoretical properties. Section \ref{section4} introduces the ACP algorithm and its feedback-driven hypothesis reweighting mechanism. Section \ref{section5} evaluates the proposed methods through numerical simulations and discusses the resulting design insights. Finally, Section \ref{section6} concludes the paper.

\textit{Notation:}
Scalars, vectors, and matrices are denoted by italic letters, bold lowercase letters, and bold uppercase letters, respectively. The sets of real and complex numbers are denoted by $\mathbb{R}$ and $\mathbb{C}$, respectively. The superscripts $(\cdot)^{\mathsf{T}}$ and $(\cdot)^{\mathsf{H}}$ denote transpose and Hermitian transpose, respectively. The expectation operator is denoted by $\mathbb{E}[\cdot]$. For a complex scalar, $\vert \cdot \vert$ and $\angle(\cdot)$ denote its modulus and phase, respectively. For a set $\mathcal{B}$, $|\mathcal{B}|$ denotes its cardinality. $(x)_+$ is defined as ${\rm max}\{x, 0\}$. Finally, $\mathcal{CN}(\mu, \sigma^2)$ denotes the circularly symmetric complex Gaussian distribution with mean $\mu$ and variance $\sigma^2$.
%%%%%%%%%%%%%%%%%%%%%%%%%%%%%%%%%%%%%%%%%%%%%%%%%%%%%%%%%%%%%%%%%%%%%%%%%%%%%%%%%%%%%%%

\begin{figure*}[t]
	\centering
	\includegraphics[width=6.1in]{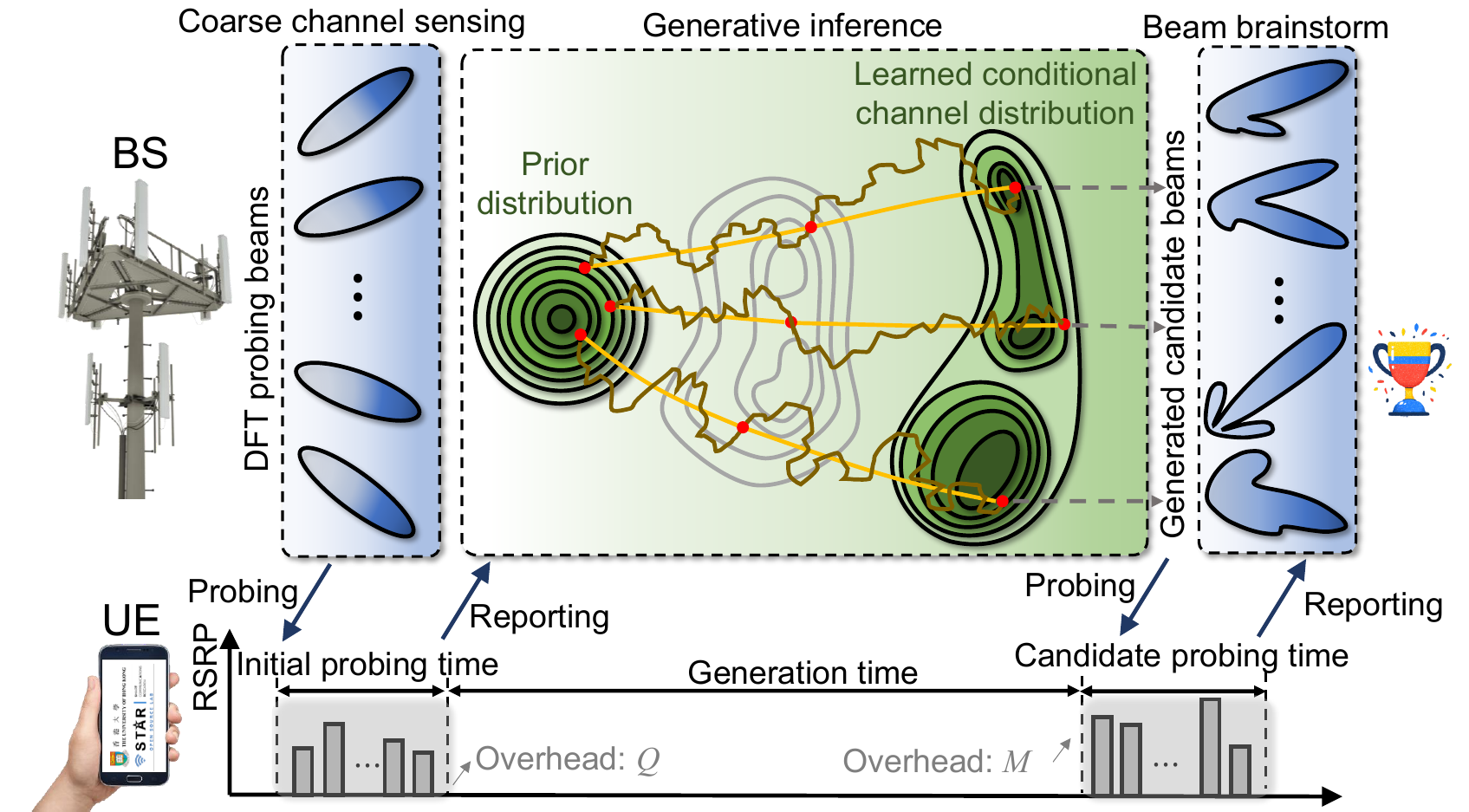}
	\caption{General procedure of GenSSBF.}
	\label{fig:beam_brainstorm}
\end{figure*}

\section{System Model and Problem Formulation}\label{section2}
Consider a downlink communication system in which a BS equipped with an $N_t$-element uniform linear array (ULA) serves a scheduled single-antenna user equipment (UE). Following \cite{zhou2026fastbeambrainstorm}, the channel between the BS and UE can be modeled as
\begin{equation}
\setlength\abovedisplayskip{3pt}%shrink space
\setlength\belowdisplayskip{3pt}
\left\{
\begin{array}{lr}
	\bm{{\rm h}} = \sum_{l=1}^L \alpha_{l}\bm{{\rm a}}(\phi_{l}), \\
	\bm{{\rm a}}(\phi_{l}) = \frac{1}{\sqrt{N_t}}\left[1, {\rm e}^{j\frac{2\pi d}{\lambda}{\sin \phi_{l}}}, \cdots, {\rm e}^{j(N_t-1)\frac{2\pi d}{\lambda}{\sin \phi_{l}}}\right]^{\mathsf{T}}.
\end{array}
\right.
\end{equation}
where $L$ denotes the number of propagation paths, $\alpha_{l}$ and $\phi_{l}$ denote the complex gain and angle of departure (AoD) of the $l$-th path between the BS and the UE, respectively, while $d$ and $\lambda$ are the antenna spacing and carrier wavelength. The BS is assumed to employ analog beamforming using phase shifters. Thus, the feasible set of beamforming vectors is
\begin{equation}\label{eq:constant_modulus_constraint}
	\setlength\abovedisplayskip{3pt}%shrink space
	\setlength\belowdisplayskip{3pt}
	\mathcal{S} \triangleq \left\{\bm{{\rm w}}\in \mathbb{C}^{N_t \times 1}: \vert w_n \vert = \frac{1}{\sqrt{N_t}}, \forall n\right\}.
\end{equation}
For given channel $\bm{{\rm h}}$, the beamforming gain achieved by $\bm{{\rm w}}$ is $g(\bm{{\rm w}}, \bm{{\rm h}}) = \vert \bm{{\rm h}}^{\mathsf{H}}\bm{{\rm w}} \vert^2$. Let $\bm{{\rm w}}^\star \in \argmax_{\bm{{\rm w}}\in \mathcal{S}} g(\bm{{\rm w}}, \bm{{\rm h}})$ denote the optimal beamforming vector when the non-zero channel $\bm{{\rm h}}$ is known. Under the constant modulus constraint in (\ref{eq:constant_modulus_constraint}), the maximum beamforming gain is achieved by aligning the phase of each antenna element with that of $\bm{{\rm h}}$, yielding the closed-form optimal beam $\bm{{\rm w}}^\star = \frac{1}{\sqrt{N_t}} {\rm e}^{j\angle \bm{{\rm h}}}$. The normalized beamforming gain is expressed as 
\begin{equation}
	\setlength\abovedisplayskip{3pt}%shrink space
	\setlength\belowdisplayskip{3pt}
	\widetilde{g}(\bm{{\rm w}}, \bm{{\rm h}}) = \frac{g(\bm{{\rm w}}, \bm{{\rm h}})}{g(\bm{{\rm w}}^\star, \bm{{\rm h}})} \in [0,1].
\end{equation}
For GenSSBF, in the initial probing stage, as illustrated in Fig.~\ref{fig:beam_brainstorm}, the BS uses $Q$ predefined DFT probing beams to obtain the RSRP observations, which is given by
\begin{equation}
	\setlength\abovedisplayskip{3pt}%shrink space
	\setlength\belowdisplayskip{3pt}
	\bm{{\rm o}}_{Q} = \left[r_{1}, r_{2}, \cdots, r_{Q}\right]^{\mathsf{T}}.
\end{equation}
Let $\bm{{\rm v}}_i$ denote the $i$-th DFT probing beam. The corresponding RSRP $r_i$ is given by
\begin{equation}
	\setlength\abovedisplayskip{3pt}%shrink space
	\setlength\belowdisplayskip{3pt}
	y_{i}=\sqrt{P_T} \bm{{\rm h}}^{\mathsf{H}} \bm{{\rm v}}_i + n_{i}, \quad r_{i} = \vert y_{i} \vert^2,
\end{equation}
where $P_T$ is the transmit power, and $n_{i}\sim \mathcal{CN}(0, \sigma^2)$ is the additive white Gaussian noise with variance $\sigma^2$ at UE during the $i$-th probing transmission. Once the UE reports the raw RSRP observations $\bm{{\rm o}}_Q$ to the BS, they are pre-processed and encoded into a conditioning representation, referred to as the wireless prompt $\bm{{\rm c}}$ \cite{zhou2026beam, zhou2026fastbeambrainstorm}. Following the coarse channel sensing stage, GenSSBF proceeds to generative inference and beam-brainstorming, as illustrated in Fig. \ref{fig:beam_brainstorm}. Specifically, a conditional generative model parameterized by $\vartheta$ is trained to learn an approximation $p_\vartheta(\bm{{\rm h}} | \bm{{\rm c}})$ to the true conditional channel distribution $p(\bm{{\rm h}} | \bm{{\rm c}})$, realized via a generator function $\mathcal{G}_\vartheta(\cdot, \cdot)$ that transforms random latent variables into plausible channel realizations. During online inference, the BS draws $M \geq 1$ latent vectors $\{\bm{{\rm z}}_m\}_{m=1}^M$ from a prescribed prior distribution $p_{\bm{{\rm z}}}$ (e.g., standard Gaussian) and feeds them, together with $\bm{{\rm c}}$, into $\mathcal{G}_\vartheta(\cdot, \cdot)$ to generate $M$ channel hypotheses $\{ \hat{\bm{{\rm h}}}_m \}_{m=1}^M$. Applying the phase-alignment mapping to each channel hypothesis $\hat{\bm{{\rm h}}}_m$ yields a corresponding constant-modulus candidate beam $\bm{{\rm w}}_m$ as
\begin{equation}\label{eq: beam_recover}
	\setlength\abovedisplayskip{3pt}%shrink space
	\setlength\belowdisplayskip{3pt}
	\hat{\bm{{\rm h}}}_m = \mathcal{G}_{\vartheta}(\bm{{\rm c}}, \bm{{\rm z}}_{m}), \quad \bm{{\rm w}}_m = \frac{1}{\sqrt{N_t}} {\rm e}^{j \angle \hat{\bm{{\rm h}}}_m} \in \mathcal{S}, m=1,\cdots,M,
\end{equation}
where $\hat{\bm{{\rm h}}}_m$ is the $m$-th generated channel hypothesis. Subsequently, the BS transmits the reference signals using these $M$ candidate beams, while the UE measures and reports the corresponding RSRPs. Finally, the BS selects the candidate beam with the highest reported RSRP for data transmission, incurring a total online probing overhead of $Q+M$.

\begin{figure}[t]
	\centering
	\includegraphics[width=3.5in]{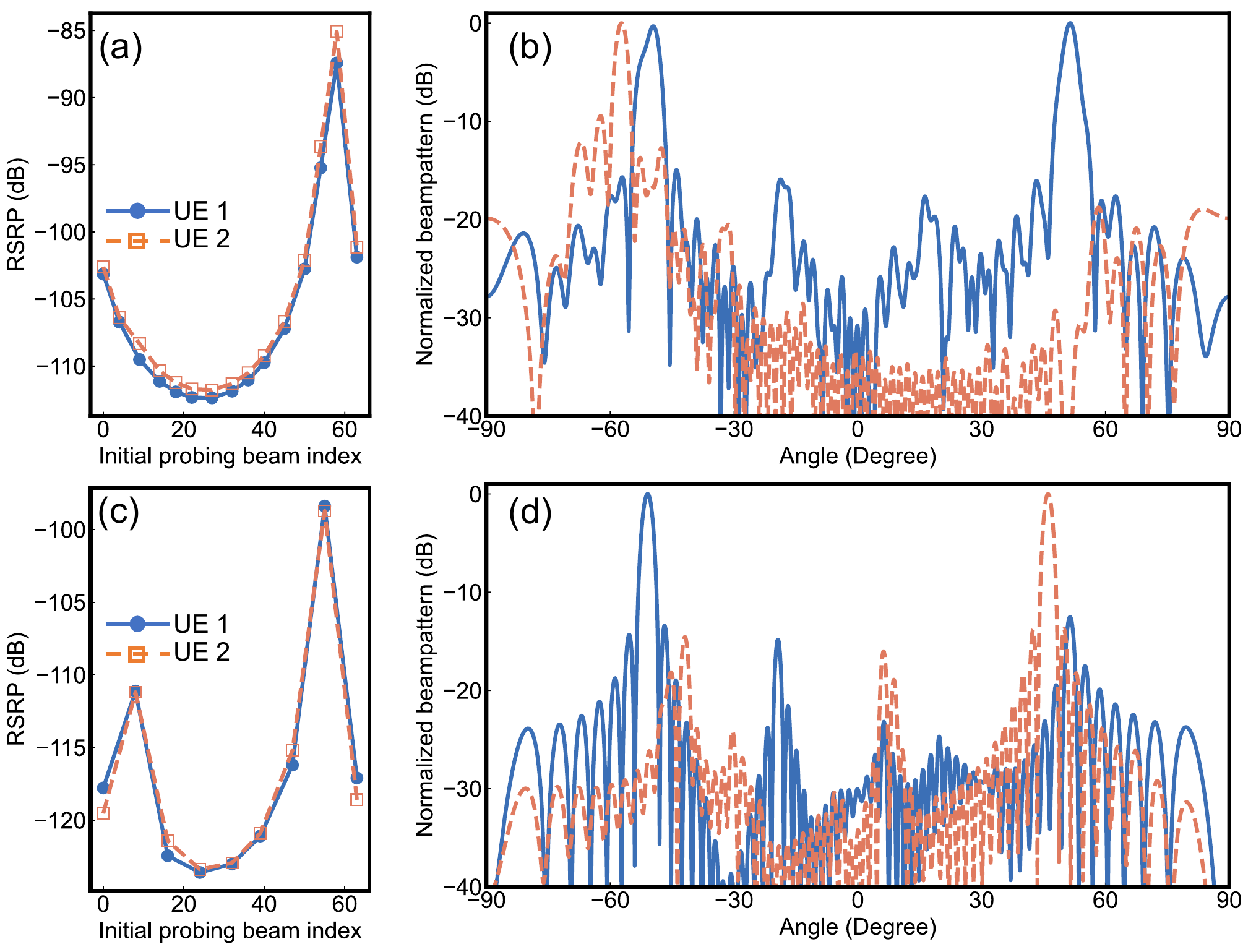}
	\caption{Similar initial RSRP observations with distinct optimal beam patterns in DeepMIMO I2\_28B: (a), (b) the first UE pair with $Q=15$; (c), (d) the second UE pair with $Q=9$.}
	\label{fig:ambiguity}
\end{figure}

Before proceeding, we first explain why beam brainstorming is needed. The wireless prompt $\bm{{\rm c}}$ provides only partial and coarse information about the underlying channel. In particular, distinct channel realizations may produce similar RSRP observations under the initial probing beams while favoring substantially different beamforming vectors. Consequently, a single generated channel hypothesis, even if plausible given $\bm{{\rm c}}$, may lead to a beam that is poorly matched to the actual channel. Fig.~\ref{fig:ambiguity} illustrates this ambiguity using two UE pairs from the DeepMIMO I2\_28B dataset \cite{alkhateeb2019deepmimo}, with $Q=15$ and $Q=9$, respectively. Within each pair, the UEs exhibit similar initial RSRP profiles, as shown in Figs.~\ref{fig:ambiguity}(a) and \ref{fig:ambiguity}(c). However, the corresponding optimal beams exhibit substantially different radiation patterns, as shown in Figs.~\ref{fig:ambiguity}(b) and \ref{fig:ambiguity}(d). Applying UE 1's optimal beam to UE 2's channel yields normalized beamforming gains of only $0.0089$ and $0.0018$ in the two cases, respectively. These examples motivate beam brainstorming, i.e., generating multiple candidate beams and evaluating them through additional online probing to identify a beam suited to the actual channel.

In the existing implementation of GenSSBF, independent latent draws produce i.i.d. channel hypotheses,
\begin{equation}\label{eq:iid_sampling}
	\setlength\abovedisplayskip{3pt}%shrink space
	\setlength\belowdisplayskip{3pt}
	\hat{\bm{{\rm h}}}_{1}, \hat{\bm{{\rm h}}}_{2}, \cdots, \hat{\bm{{\rm h}}}_{M} \stackrel{{\rm i.i.d.}}{\sim} p_\vartheta (\cdot | \bm{{\rm c}}),
\end{equation}
which are mapped to candidate beams through (\ref{eq: beam_recover}). As mentioned before, all these $M$ candidate beams need to be probed, resulting in a total online probing overhead of $Q+M$. Although additional candidates provide more opportunities to identify a suitable beam, each verified candidate consumes an online probing opportunity. Under a limited probing budget, \textit{it is therefore necessary to consider whether a candidate beam is worth probing}\footnote{This question is also relevant to multi-user extensions. For instance, the two-stage hybrid beamforming framework in \cite{alkhateeb2015limited} first designs analog beams separately for each UE and then applies digital beamforming to manage inter-user interference. Embedding GenSSBF in this framework would require beam brainstorming for each UE in the first stage, making efficient candidate probing particularly important.}. In particular, independently generated candidates may provide overlapping gains across plausible channel realizations, so an additional candidate may offer little improvement over those already included in the set. This raises the following question: \textit{which candidates should be probed to maximize the expected gain of the finally selected beam?}

Therefore, our goal is to design a candidate beam set that makes effective use of each probing opportunity under the channel uncertainty remaining after initial coarse sensing. Let set $\mathcal{B} \subseteq \mathcal{S}$ denote a finite set of candidate beams. For a given channel $\bm{{\rm h}}$, the maximum normalized beamforming gain achievable by the beams in $\mathcal{B}$ is defined as
\begin{equation}\label{eq:widetilde_g}
	\setlength\abovedisplayskip{3pt}%shrink space
	\setlength\belowdisplayskip{3pt}
	\widetilde{g}(\mathcal{B}, \bm{{\rm h}}) \triangleq \max_{\bm{{\rm w}} \in \mathcal{B}} \widetilde{g}(\bm{{\rm w}}, \bm{{\rm h}}), \quad \widetilde{g}(\emptyset, \bm{{\rm h}}) = 0.
\end{equation}
Since the actual channel $\bm{{\rm h}}$ is unknown and follows the conditional distribution $p(\bm{{\rm h}} | \bm{{\rm c}})$, we evaluate the effectiveness of a candidate set $\mathcal{B}$ by its expected normalized beamforming gain, which is given by
\begin{equation}\label{eq:objective}
	\setlength\abovedisplayskip{3pt}%shrink space
	\setlength\belowdisplayskip{3pt}
	V(\mathcal{B}) \triangleq \mathbb{E}_{p(\bm{{\rm h}} | \bm{{\rm c}})}\left[\widetilde{g}(\mathcal{B}, \bm{{\rm h}})\right] = \mathbb{E}_{p(\bm{{\rm h}} | \bm{{\rm c}})}\left[\max_{\bm{{\rm w}} \in \mathcal{B}} \widetilde{g}(\bm{{\rm w}}, \bm{{\rm h}})\right].
\end{equation}
Under a probing budget of $M$ candidate beams, the candidate set design problem is formulated as
\begin{subequations}
	\setlength\abovedisplayskip{3pt}%shrink space
	\setlength\belowdisplayskip{3pt}
	\begin{align}
		(\textbf{P1}): \max_{\mathcal{B} \subset \mathcal{S}} \quad & V(\mathcal{B}) \\
		\mathrm{s.t.} \quad & |\mathcal{B} | \leq M. \label{constraint1}
	\end{align}
\end{subequations}
Specifically, for each plausible channel realization $\bm{{\rm h}}$, the inner maximization $\max_{\bm{{\rm w}} \in \mathcal{B}} \widetilde{g}(\bm{{\rm w}}, \bm{{\rm h}})$ quantifies the highest beamforming gain attainable within $\mathcal{B}$, which corresponds to the gain achieved under ideal beam selection. The outer expectation over $p(\bm{{\rm h}} | \bm{{\rm c}})$ averages this achievable gain across all plausible channels consistent with the wireless prompt $\bm{{\rm c}}$, thereby accounting for the residual channel ambiguity. Moreover, adopting the normalized gain can prevent channel realizations with larger path gains from dominating the expectation. However, directly solving (\textbf{P1}) is challenging because the true conditional channel distribution is generally unavailable, and the candidate beams are coupled in the objective function. Section \ref{section3} presents a greedy approximation to (\textbf{P1}). Section \ref{section4} extends this approach by using RSRP feedback from earlier candidate probes to guide subsequent selections.

\section{Greedy Candidate Probing (GCP)}\label{section3}
For (\textbf{P1}), one of the most straightforward feasible solutions is to construct candidate beams from independently generated channel hypotheses, as shown in (\ref{eq: beam_recover}) and (\ref{eq:iid_sampling}). However, this solution is not necessarily a ``resource-efficient'' way since independently constructed beams may perform well on the same channel realizations. It means that once one of these beams is included in the candidate set, adding the others may provide little additional gain. Under a limited probing budget, each additional candidate should therefore be chosen based on \textit{how much it improves upon the beams already selected}. Independent sampling does not explicitly account for this dependence, so probing more candidates may consume additional resources without substantially improving the beamforming performance of the system.

\begin{tcolorbox}[
	title={Illustrative example},
	colback=white, colframe=black,
	colbacktitle=white!80!gray, coltitle=black, 
	breakable,
	float=!t,
	% float,floatplacement=!b,
	before upper={\parindent10pt\noindent},  
	left = 1mm, 
	right = 1mm,
	top = 1mm,
	bottom = 1mm
	]

	Consider two possible channel realizations with posterior probabilities 0.8 and 0.2. Suppose that the current set $\mathcal{B}$ achieves normalized beamforming gain of 0.9 and 0.1, respectively, yielding $V(\mathcal{B})=0.8\times 0.9 + 0.2 \times 0.1 = 0.74$. Two additional candidates, $\bm{{\rm w}}_A$ and $\bm{{\rm w}}_B$, provide gains of $(0.95, 0.05)$ and $(0.20, 0.80)$. Their standalone expected normalized gains are $0.8\times 0.95 + 0.2\times 0.05 = 0.77$ and $0.8\times 0.2 + 0.2\times 0.8 = 0.32$, respectively. Nevertheless, their marginal probing values satisfy
	\begin{align}
		\setlength\abovedisplayskip{3pt}%shrink space
		\setlength\belowdisplayskip{3pt}
		\Delta(\bm{{\rm w}}_A | \mathcal{B}) & = 0.8\times (0.95 - 0.9)_+ \nonumber \\
		& + 0.2\times (0.05 - 0.1)_+ = 0.04,  \nonumber
	\end{align}
	and
	\begin{align}
		\setlength\abovedisplayskip{3pt}%shrink space
		\setlength\belowdisplayskip{3pt}
		\Delta(\bm{{\rm w}}_B | \mathcal{B}) & = 0.8\times (0.2 - 0.9)_+ \nonumber \\
		& + 0.2\times (0.8 - 0.1)_+ = 0.14. \nonumber
	\end{align}
	resulting in 
	\begin{equation}
		\setlength\abovedisplayskip{3pt}%shrink space
		\setlength\belowdisplayskip{3pt}
		V(\mathcal{B} \cup \{ \bm{{\rm w}}_A \}) = 0.78 < V(\mathcal{B} \cup \{ \bm{{\rm w}}_B \}) = 0.88. \nonumber
	\end{equation}
	Although $\bm{{\rm w}}_A$ performs better in isolation, $\bm{{\rm w}}_B$ yields a larger improvement to the current set by serving the channel realization on which $\mathcal{B}$ performs poorly. This example illustrates why an additional probing opportunity should be allocated according to a candidate’s marginal contribution rather than its standalone expected gain.
\end{tcolorbox}

To evaluate the performance gain from introducing an additional candidate beam $\bm{{\rm w}} \in \mathcal{S}$ to an existing candidate set $\mathcal{B}\subseteq \mathcal{S}$, we define the \textit{marginal probing value} based on the objective $V(\mathcal{B})$ in (\ref{eq:objective}), which is given by
\begin{subequations}
	\setlength\abovedisplayskip{3pt}%shrink space
	\setlength\belowdisplayskip{3pt}
	\begin{align}
		\Delta(\bm{{\rm w}} | \mathcal{B}) & \triangleq V(\mathcal{B} \cup \{\bm{{\rm w}}\}) - V(\mathcal{B}) \\
		& = \mathbb{E}_{p(\bm{{\rm h}} | \bm{{\rm c}})}\left[(\widetilde{g}(\bm{{\rm w}}, \bm{{\rm h}}) - \widetilde{g}(\mathcal{B}, \bm{{\rm h}}))_+\right] \label{eq:14b}\\
		& =\mathbb{E}_{p(\bm{{\rm h}} | \bm{{\rm c}})}\left[\widetilde{g}(\bm{{\rm w}}, \bm{{\rm h}})\right] \!\!-\!\! \mathbb{E}_{p(\bm{{\rm h}} | \bm{{\rm c}})}\left[\min \{\widetilde{g}(\bm{{\rm w}}, \bm{{\rm h}}), \widetilde{g}(\mathcal{B},\! \bm{{\rm h}})\}\right], \label{eq:marginal_value}
	\end{align}
\end{subequations}
where $(x)_+ = \max\{x, 0\}$, and the last equality follows from the identity $(a-b)_+ = a-\min\{a, b\}$. The first term in (\ref{eq:marginal_value}) represents the standalone expected beamforming gain of $\bm{{\rm w}}$, while the second term quantifies the expected gain overlap between $\bm{{\rm w}}$ and the existing candidate set $\mathcal{B}$. Specifically, for any channel realization $\bm{{\rm h}}$, $\min \{\widetilde{g}(\bm{{\rm w}}, \bm{{\rm h}}), \widetilde{g}(\mathcal{B},\! \bm{{\rm h}})\}$ equals the portion of $\widetilde{g}(\bm{{\rm w}}, \bm{{\rm h}})$ that is already covered by $\mathcal{B}$. For instance, if $\widetilde{g}(\bm{{\rm w}}, \bm{{\rm h}}) \leq \widetilde{g}(\mathcal{B},\! \bm{{\rm h}})$, the entire gain of $\bm{{\rm w}}$ on channel $\bm{{\rm h}}$ is redundant and yields zero marginal improvement. Otherwise, only the excess gain beyond $\widetilde{g}(\mathcal{B}, \bm{{\rm h}})$ contributes to the set gain. Thus, the decomposition in (\ref{eq:marginal_value}) shows that the standalone expected beamforming gain is insufficient to determine whether it is worth an additional online probe. A candidate beam may have a high beamforming gain on its own but offer little additional benefit because of its overlap with the selected beams. In this sense, the marginal value is ``\textit{set-dependent}''. As illustrated in the example box, a candidate with a lower standalone expected gain can provide a larger marginal probing value by improving performance on channel realizations poorly served by the selected beams.

The marginal value defined in (\ref{eq:marginal_value}) provides a natural criterion for constructing the candidate set. Instead of determining the beams independently, the BS can actively select each candidate according to its additional contribution to the current set. This leads to a greedy approximation to (\textbf{P1}). Specifically, let $\mathcal{B}_0 = \emptyset$. At iteration $m$, the ideal greedy procedure solves
\begin{subequations}
	\setlength\abovedisplayskip{3pt}%shrink space
	\setlength\belowdisplayskip{3pt}
	\begin{align}\label{eq:15a}
		(\textbf{P2}): \bm{{\rm w}}_m \in \argmax_{\bm{{\rm w}} \in \mathcal{S} \backslash \mathcal{B}_{m-1}} \Delta(\bm{{\rm w}} | \mathcal{B}_{m-1}),
	\end{align}
\end{subequations}
and updates $\mathcal{B}_m = \mathcal{B}_{m-1} \cup \{\bm{{\rm w}}_m\}, m=1,\cdots, M$.

However, directly solving (\textbf{P2}) presents two difficulties. First, evaluating $\Delta$ requires an expectation under the true conditional channel distribution, which is generally unavailable. Second, maximizing this expectation over the continuous feasible set $\mathcal{S}$ is computationally challenging. Therefore, in the following, we address these difficulties by developing a plug-and-play greedy candidate probing method based on generated channel hypotheses.

\begin{figure*}[t]
	\centering
	\includegraphics[width=7.2in]{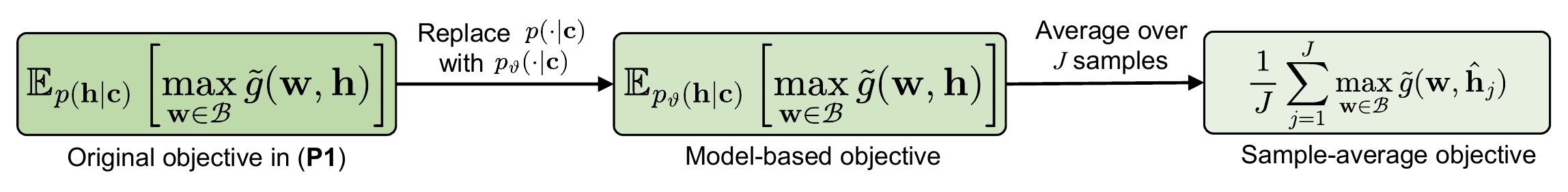}
	\caption{Two-step approximation of the objective in (\textbf{P1}) used by GCP. The true conditional channel distribution is replaced by the learned distribution, and the resulting expectation is estimated by averaging over $J$ generated channel hypotheses.}
	\label{fig:approximation_in_GCP}
\end{figure*}

\subsection{Plug-and-Play Greedy Candidate Selection}\label{section3.A}
To overcome these obstacles, we develop a plug-and-play greedy candidate probing (GCP) algorithm that provides a practical approximation to (\textbf{P2}). First, GCP draws a finite set of channel hypotheses from the pre-trained generative model and uses their sample average to approximate the expectation in (\ref{eq:15a}). At the same time, these channel hypotheses also yield feasible candidate beams, allowing the continuous search in (\textbf{P2}) to be restricted to a finite pool. Specifically, given the wireless prompt $\bm{{\rm c}}$, we generate $J \geq M$ channel hypotheses as in (\ref{eq:iid_sampling}). For each nonzero hypothesis, the normalized beamforming gain of a feasible beam $\bm{{\rm w}}$ is
\begin{equation}
	\setlength\abovedisplayskip{3pt}%shrink space
	\setlength\belowdisplayskip{3pt}
	\widetilde{g}(\bm{{\rm w}}, \hat{\bm{{\rm h}}}_j) = \frac{\vert \hat{\bm{{\rm h}}}_j^\mathsf{H} \bm{{\rm w}} \vert^2}{\max_{\bm{{\rm w}}' \in \mathcal{S}} \vert \hat{\bm{{\rm h}}}_j^\mathsf{H} \bm{{\rm w}}' \vert^2} = \frac{\vert \hat{\bm{{\rm h}}}_j^\mathsf{H} \bm{{\rm w}} \vert^2}{\frac{1}{N_t}\left(\sum_{n=1}^{N_t} \vert \hat{h}_{j,n} \vert  \right)^2}.
\end{equation}
In order to obtain a tractable surrogate for the expected normalized beamforming gain $V(\mathcal{B})$ in (\textbf{P1}), we adopt a two-step approximation, as illustrated in Fig. \ref{fig:approximation_in_GCP}. First, since the true conditional channel distribution $p(\bm{{\rm h}} | \bm{{\rm c}})$ is unavailable, we replace it with the learned distribution $p_\vartheta(\bm{{\rm h}} | \bm{{\rm c}})$. Second, we approximate the resulting model-based expectation by an equally weighted sample average over the $J$ i.i.d. channel hypotheses drawn from $p_\vartheta(\bm{{\rm h}} | \bm{{\rm c}})$, where each channel hypothesis carries a weight of $1/J$. This two-step approximation yields the following sample-average objective:
\begin{equation}\label{eq:estimated_V}
	\setlength\abovedisplayskip{3pt}%shrink space
	\setlength\belowdisplayskip{3pt}
	\hat{V}(\mathcal{B}) = \frac{1}{J}\sum_{j=1}^J \max_{\bm{{\rm w}} \in \mathcal{B}} \widetilde{g}(\bm{{\rm w}}, \hat{\bm{{\rm h}}}_j), \quad \hat{V}(\emptyset) = 0.
\end{equation}
Note that the accuracy of $\hat{V}(\mathcal{B})$ in (\ref{eq:estimated_V}) is governed by two factors corresponding to these two steps. First, as $J\to \infty$, the sample average converges to the expectation under $p_\vartheta(\bm{{\rm h}} | \bm{{\rm c}})$, so increasing $J$ reduces sampling variability. However, any bias arising from the distribution mismatch between $p(\bm{{\rm h}} | \bm{{\rm c}})$ and $p_\vartheta(\bm{{\rm h}} | \bm{{\rm c}})$ persists even when the candidate pool is very large.
 
Based on (\ref{eq:estimated_V}), the estimated marginal probing value is
\begin{align}\label{eq:estimated_posterior_expectation}
	\setlength\abovedisplayskip{3pt}%shrink space
	\setlength\belowdisplayskip{3pt}
	\hat{\Delta}(\bm{{\rm w}} | \mathcal{B}) &= \hat{V}(\mathcal{B} \cup \{\bm{{\rm w}}\}) - \hat{V}(\mathcal{B}) \nonumber \\
	&=\frac{1}{J}\sum_{j=1}^J \left(\widetilde{g}(\bm{{\rm w}}, \hat{\bm{{\rm h}}}_j) - \widetilde{g}(\mathcal{B}, \hat{\bm{{\rm h}}}_j)\right)_+.
\end{align}
As mentioned before, (\ref{eq:estimated_posterior_expectation}) evaluates a candidate through its remaining contribution across the generated hypotheses, a hypothesis contributes to the score only when the candidate improves upon the best beamforming gain already provided by $\mathcal{B}$.
 
This approximation makes the marginal value computable for a given beam, but maximizing it over the continuous feasible set $\mathcal{S}$ remains challenging. We therefore construct a finite candidate pool by recovering one beam from each of the $J$ hypotheses using (\ref{eq: beam_recover}):
\begin{equation}
	\setlength\abovedisplayskip{3pt}%shrink space
	\setlength\belowdisplayskip{3pt}
	\bm{{\rm w}}_{j} = \frac{1}{\sqrt{N_t}}\left[{\rm e}^{j\angle \hat{h}_{j,1}}, \cdots, {\rm e}^{j\angle \hat{h}_{j,N_t}}\right]^{\mathsf{T}}, j=1,\cdots,J.
\end{equation}
The resulting candidate pool is denoted by $\mathcal{C}_{J}\triangleq \{\bm{{\rm w}}_{1}, \cdots, \bm{{\rm w}}_{J}\}$. Thus, each generated sample provides both a channel hypothesis for evaluating the expected value and a feasible beam for candidate selection. To evaluate how each beam serves every channel hypothesis, define the normalized gain matrix as
\begin{equation}\label{eq:gain_matrix}
	\setlength\abovedisplayskip{3pt}%shrink space
	\setlength\belowdisplayskip{3pt}
	\bm{{\rm G}} \in \mathbb{R}^{J\times J}_+, \quad G_{i,j} \triangleq \widetilde{g}(\bm{{\rm w}}_i, \hat{\bm{{\rm h}}}_j).
\end{equation}
The $i$-th row describes the normalized beamforming gains achieved by candidate $i$ across all channel hypotheses, while the $j$-th column describes how the candidates perform on channel hypothesis $j$. Since $\bm{{\rm w}}_j$ maximizes the constant-modulus beamforming gain for $\hat{\bm{{\rm h}}}_j$, the diagonal entries satisfy $G_{j,j}=1$. The off-diagonal entries quantify how well each candidate also serves the other hypotheses. Let $\mathcal{I}_{m-1}$ denote the indices selected before iteration $m$, with corresponding beam set $\mathcal{B}_{m-1} = \{\bm{{\rm w}}_i: i\in \mathcal{I}_{m-1}\}$. The best gain currently provided on hypothesis $j$ is
\begin{equation}
	\setlength\abovedisplayskip{3pt}%shrink space
	\setlength\belowdisplayskip{3pt}
	a_{j}^{(m-1)} = \left\{
	\begin{aligned}
		& \max_{i\in \mathcal{I}_{m-1}} G_{i,j}, \quad & \mathcal{I}_{m-1} \neq \emptyset, \\
		& 0, \quad & \mathcal{I}_{m-1} = \emptyset.
	\end{aligned}
	\right.
\end{equation}
Replacing the marginal probing value in (\textbf{P2}) with its estimate in (\ref{eq:estimated_posterior_expectation}) and restricting the search to the unselected candidates yields
\begin{subequations}
	\setlength\abovedisplayskip{3pt}%shrink space
	\setlength\belowdisplayskip{3pt}
	\begin{align}
		(\textbf{P3}): i_m \in \argmax_{i\in\{1,\cdots,J\} \backslash \mathcal{I}_{m-1}} \frac{1}{J}\sum_{j=1}^J\left(G_{i,j} - a_{j}^{(m-1)}\right)_+. \label{eq: greedy_marginal_probing_value}
	\end{align}
\end{subequations}
After selecting $i_m$, the BS updates $\mathcal{I}_m = \mathcal{I}_{m-1} \cup \{i_m\}$, $\mathcal{B}_m = \mathcal{B}_{m-1} \cup \{\bm{{\rm w}}_{i_m}\}$, and $a_{j}^{(m)} = \max \{a_{j}^{(m-1)}, G_{i_m, j}\}, j=1,\cdots,J$. The procedure is initialized with $\mathcal{I}_0 = \emptyset$ and $a_{j}^{(0)}=0$, and repeated for $M$ iterations. Although the candidate pool itself is generated through i.i.d. sampling, the selected candidates are jointly determined through their marginal contributions, independent generation provides a computationally accessible pool, and greedy selection determines which members of that pool merit online probing. The procedure of GCP is summarized in \textbf{Algorithm 1}.

\begin{algorithm}[t]
	\caption{Greedy Candidate Probing (GCP)}
	\label{alg:greedy_selection}
	\begin{algorithmic}[1]
		\REQUIRE Wireless prompt $\mathbf c$, trained generator
		$\mathcal G_{\vartheta}$, pool size $J$, and probing
		budget $M \leq J$.
		\ENSURE Candidate set $\mathcal B_M$ and communication beam
		$\mathbf w_{\mathrm{com}}$.
		
		\STATE Generate $J$ independent channel hypotheses
		$\{\widehat{\mathbf h}_j\}_{j=1}^{J}$ given $\mathbf c$.
		Recover beam $\mathbf w_j$ from each hypothesis
		via \eqref{eq: beam_recover}.
		\STATE Form the pool $\mathcal C_J=\{\mathbf w_j\}_{j=1}^{J}$
		and compute
		$G_{i,j}=\widetilde g(\mathbf w_i,\widehat{\mathbf h}_j)$
		for all $i,j$.
		\STATE Set $\mathcal I_0=\emptyset$ and $a_j^{(0)}=0$ for all $j$.
		
		\FOR{$m=1,\ldots,M$}
		\FORALL{$i\in\{1,\ldots,J\}\setminus\mathcal I_{m-1}$}
		\STATE $\displaystyle
		d_i \gets \frac{1}{J}\sum_{j=1}^{J}
		\bigl(G_{i,j}-a_j^{(m-1)}\bigr)_+$
		\ENDFOR
		\STATE Choose $i_m\in
		\argmax_{i\notin\mathcal I_{m-1}} d_i$.
		\STATE $\mathcal I_m\gets\mathcal I_{m-1}\cup\{i_m\}$
		\STATE $a_j^{(m)}\gets
		\max\{a_j^{(m-1)},G_{i_m,j}\}$ for all $j$.
		\ENDFOR
		
		\STATE Form $\mathcal B_M=\{\mathbf w_{i_m}\}_{m=1}^{M}$.
		\STATE Probe the beams in $\mathcal B_M$ and obtain their
		reported RSRPs $\{r_m\}_{m=1}^{M}$.
		\STATE Choose $m^\star\in\argmax_{1\leq m\leq M}r_m$
		and set $\mathbf w_{\mathrm{com}}=\mathbf w_{i_{m^\star}}$.
	\end{algorithmic}
\end{algorithm}

\subsection{Performance Guarantees and Computational Complexity}
The proposed method approximates (\textbf{P1}) by constructing a finite candidate pool, estimating posterior expected values from generated hypotheses, and selecting candidates greedily. This subsection establishes the structural properties of the proposed GCP, and provides the computational complexity analysis. Throughout the analysis, the wireless prompt $\bm{{\rm c}}$ and the parameters $\vartheta$ of the trained generative model are fixed. The generated hypotheses $\hat{\bm{{\rm h}}}_1, \cdots, \hat{\bm{{\rm h}}}_J$ are assumed to be independent samples from $p_\vartheta(\cdot | \bm{{\rm c}})$, and each $\bm{{\rm w}}_j$ is recovered from its corresponding hypothesis through the fixed mapping in (\ref{eq: beam_recover}). All channel realizations considered in the normalized gain definitions are nonzero.

\subsubsection{Properties and Approximation Guarantees}
To establish an approximation guarantee for GCP, we first characterize the structural properties of the objectives.
\begin{proposition}\label{proposition1}
	The set functions $V$ and $\hat{V}$ for every fixed realization of the generated samples, are non-negative, monotone, and submodular, and take value zero at the empty set.
\end{proposition}

\begin{IEEEproof}
	For any finite sets $\mathcal{A} \subset \mathcal{B} \subset \mathcal{S}$ and any channel realization $\bm{{\rm h}}$, according to the definition of (\ref{eq:widetilde_g}), we have
	\begin{equation}
		\setlength\abovedisplayskip{3pt}%shrink space
		\setlength\belowdisplayskip{3pt}
		\widetilde{g}(\mathcal{A}, \bm{{\rm h}}) \leq \widetilde{g}(\mathcal{B}, \bm{{\rm h}}).
	\end{equation}
	Consequently, for an additional beam $\bm{{\rm w}} \notin \mathcal{B}$,
	\begin{align}
		\setlength\abovedisplayskip{3pt}%shrink space
		\setlength\belowdisplayskip{3pt}
		& \widetilde{g}(\mathcal{A} \cup \{\bm{{\rm w}}\}, \bm{{\rm h}}) - \widetilde{g}(\mathcal{A}, \bm{{\rm h}}) = ( \widetilde{g}(\bm{{\rm w}}, \bm{{\rm h}}) - \widetilde{g}(\mathcal{A}, \bm{{\rm h}}) )_+ \nonumber \\
		& \geq ( \widetilde{g}(\bm{{\rm w}}, \bm{{\rm h}}) - \widetilde{g}(\mathcal{B}, \bm{{\rm h}}) )_+ = \widetilde{g}(\mathcal{B} \cup \{\bm{{\rm w}}\}, \bm{{\rm h}}) - \widetilde{g}(\mathcal{B}, \bm{{\rm h}}).
	\end{align}
	Taking an expectation under either conditional distribution, or averaging over the generated hypotheses, preserves this inequality and establishes submodularity. Nonnegativity, monotonicity, and the value at the empty set follow directly from the definition of the normalized set gain. The proof is thus completed.
\end{IEEEproof}

The above proposition characterizes the diminishing marginal value of candidate probing. Specifically, for any finite beam sets $\mathcal{A} \subset \mathcal{B} \subset \mathcal{S}$ and candidate $\bm{{\rm w}} \in \mathcal{S} \backslash \mathcal{B}$, we have
\begin{equation}
	\setlength\abovedisplayskip{3pt}%shrink space
	\setlength\belowdisplayskip{3pt}
	\hat{\Delta}(\bm{{\rm w}} | \mathcal{A}) \geq \hat{\Delta}(\bm{{\rm w}} | \mathcal{B}).
\end{equation}
The same inequality holds for the posterior marginal probing value $\Delta$. As the selected set provides higher gains across plausible channels, the additional improvement available from another candidate decreases. This property formalizes the gain overlap discussed in Section \ref{section3.A} and motivates selecting candidates according to their marginal contributions. For a fixed generated pool $\mathcal{C}_J$, under a probing budget of $M$, define the estimated optimal posterior expected value within this generated pool as
\begin{equation}
	\setlength\abovedisplayskip{3pt}%shrink space
	\setlength\belowdisplayskip{3pt}
	\hat{V}^\star_{\mathcal{C}_J, M} \triangleq \max_{\mathcal{B} \subset \mathcal{C}_J, \vert \mathcal{B} \vert \leq M} \hat{V}(\mathcal{B}).
\end{equation}
Since GCP selects the candidate with the largest estimated marginal value at each iteration, the classical guarantee for monotone submodular maximization under a cardinality constraint gives
\begin{equation}\label{eq:guarantee}
	\setlength\abovedisplayskip{3pt}%shrink space
	\setlength\belowdisplayskip{3pt}
	\hat{V}(\mathcal{B}_M) \geq \alpha_M \hat{V}^\star_{\mathcal{C}_J, M}, \quad  \alpha_M \triangleq 1- \left(1 - \frac{1}{M}\right)^M \geq 1 - \frac{1}{e},
\end{equation}
where $\mathcal{B}_M$ is the candidate set returned by GCP. This bound follows from the result of \cite{nemhauser1978analysis}, it means that the greedy set $\mathcal{B}_M$ achieves at least an $\alpha_M$ fraction of the largest sample-based value attainable by selecting $M$ candidates from $\mathcal{C}_J$. It should be noted that (\ref{eq:guarantee}) applies to the estimated value $\hat{V}$. The performance under the true posterior objective $V$ additionally depends on the quality of the generated hypotheses and the coverage of the candidate pool.

\subsubsection{Computational Complexity}
Let $C_{{\rm gen}}$ denote the computational cost of generating one channel hypothesis and its associated beamforming vector. Thus, obtaining $J$ candidate beams requires $\mathcal{O}(JC_{{\rm gen}})$ operations. Constructing the gain matrix $\bm{{\rm G}}$ in (\ref{eq:gain_matrix}) requires $\mathcal{O}(J^2N_t)$ operations, since each entry involves an inner product of length $N_t$. At iteration $m$, evaluating the marginal probing value of the $J-m+1$ remaining candidates costs $\mathcal{O}(J(J-m+1))$, while selecting the next beam and updating the covered gains require $\mathcal{O}(J)$ operations. Summing over $M$ iterations yields a selection complexity of $\mathcal{O}(MJ^2)$. Consequently, the overall computational complexity of GCP is $\mathcal{O}(JC_{{\rm gen}} + J^2N_t + MJ^2)$.

\section{Adaptive Candidate Probing (ACP)} \label{section4}
The greedy method in Section \ref{section3} constructs the entire candidate set using the initial wireless prompt $\bm{{\rm c}}$. Although its selection criterion accounts for the set dependency, it does not use the measurements obtained during candidate probing, which provide additional evidence about the underlying channel and can change which remaining candidates are worth probing. We therefore develop an adaptive procedure that alternates between candidate selection, online probing, and weight updating.

The adaptive method uses the same generated channel hypotheses and candidate pool as GCP. Its additional capability lies in revising the relative weights of the hypotheses after each probing measurement. The next candidate is then selected according to its marginal probing value under the updated belief about the channel.

\subsection{Adaptive Beam Selection}
In (\ref{eq: greedy_marginal_probing_value}), the marginal probing value of a candidate beam is estimated by averaging its gain improvement over $J$ channel hypotheses. Each hypothesis contributes a weight of $1 / J$. Once the first beam has been probed, the BS has additional information. For instance, two channel hypotheses that were both plausible under $\bm{{\rm c}}$ may predict different responses to this beam. If the RSRP measurement favors one prediction, the gain improvements available under that hypothesis should receive greater emphasis when selecting the second beam. In this work, we achieve this by updating weights of channel hypotheses to reflect the accumulated feedback.

Let $\mathcal{D}_{m-1}$ collect the probed beam indices and their measured RSRPs after $m-1$ probing steps. Given this history, the marginal probing value of an unprobed candidate beam $\bm{{\rm w}}_i$ is
\begin{equation}
	\setlength\abovedisplayskip{3pt}%shrink space
	\setlength\belowdisplayskip{3pt}
	\Delta(\bm{{\rm w}}_i | \mathcal{B}_{m-1}) = \mathbb{E}_{\bm{{\rm h}} | \bm{{\rm c}}, \mathcal{D}_{m-1}}\left[(\widetilde{g}(\bm{{\rm w}}_i, \bm{{\rm h}}) - \widetilde{g}(\mathcal{B}_{m-1}, \bm{{\rm h}}) )_+\right].
\end{equation}
The gain improvement inside the expectation is the same as in (\ref{eq:14b}), but the distribution over which it is averaged now also depends on $\mathcal{D}_{m-1}$. To approximate this updated expectation, let $\omega_j^{(m-1)} \geq 0$ denote the weight assigned to $\hat{\bm{{\rm h}}}_j$ after $m-1$ measurements, with $\sum_{j=1}^J \omega_j^{(m-1)} = 1$. Thus, the estimated marginal probing value now becomes
\begin{equation}
	\setlength\abovedisplayskip{3pt}%shrink space
	\setlength\belowdisplayskip{3pt}
	\hat{\Delta}(\bm{{\rm w}}_i | \mathcal{B}_{m-1}) = \sum_{j=1}^J \omega_j^{(m-1)} \left(G_{i,j} -a_j^{(m-1)}\right)_+.
\end{equation}
Consequently, the BS selects the next candidate beam according to 
\begin{subequations}
	\setlength\abovedisplayskip{3pt}%shrink space
	\setlength\belowdisplayskip{3pt}
	\begin{align}
		(\textbf{P4}): i_m \in \argmax_{i\in\{1,\cdots,J\} \backslash \mathcal{I}_{m-1}} \sum_{j=1}^J\omega_j^{(m-1)} \left(G_{i,j} - a_{j}^{(m-1)}\right)_+. \label{eq: adaptive_marginal_probing_value}
	\end{align}
\end{subequations}
Before the first probe, $\omega_j^{(0)} = 1 / J$, so the first candidate beam is determined based on GCP. After each probe, the measured RSRP provides the evidence for updating $\omega_j^{(m-1)}$. The following subsection describes how the BS obtains these weights from the feedback.

\begin{figure}[t]
	\centering
	\includegraphics[width=3.3in]{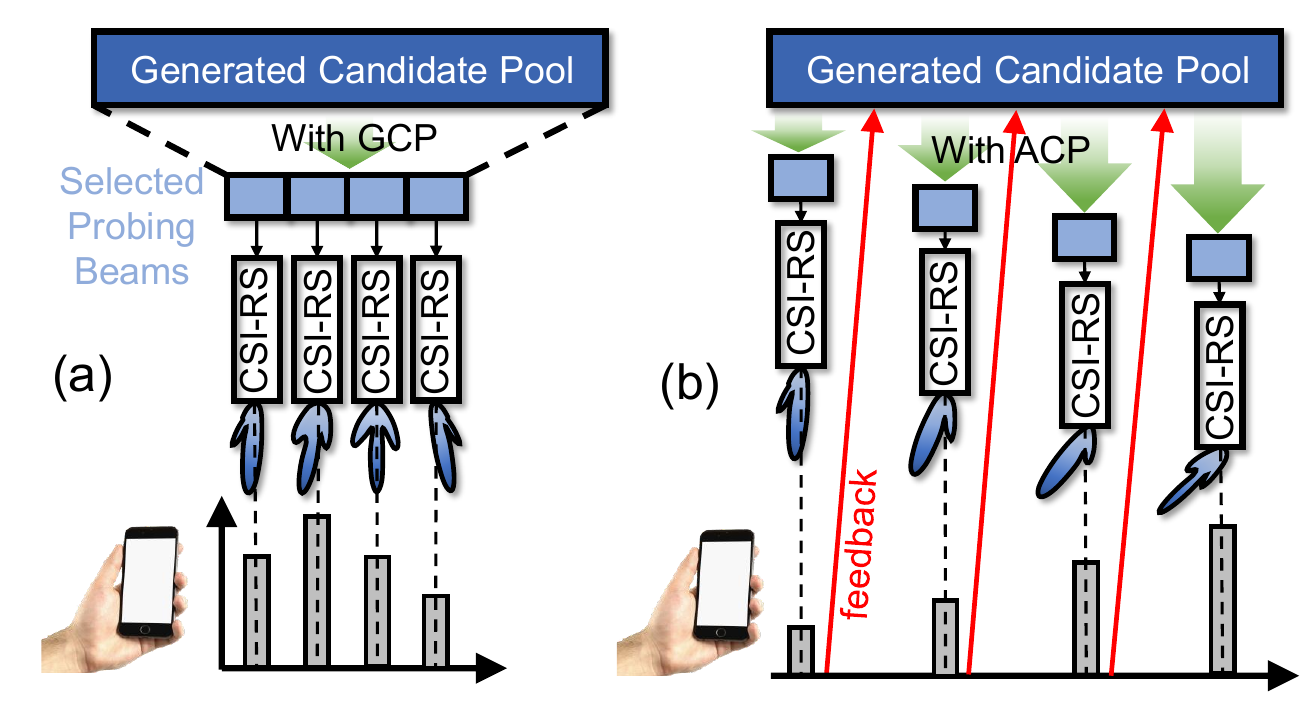}
	\caption{Candidate probing procedures: (a) GCP determines the candidate set before probing; (b) ACP uses feedback from previous probes to guide subsequent candidate selection.}
	\label{fig:gcp_and_acp}
\end{figure}

\subsection{Weight Updating from RSRP Feedback}\label{section4.B}
Suppose that the BS probes candidate beam $\bm{{\rm w}}_m$ at step $m$. For this beam, the UE reports a single RSRP measurement, while each channel hypothesis provides its own prediction of that measurement. Comparing these predictions with the reported RSRP allows the BS to assess which hypotheses are better supported by this new observation.

Let $\hat{l}_{m, j}$ denote the RSRP predicted under hypothesis $\hat{\bm{{\rm h}}}_j$ for beam $\bm{{\rm w}}_m$ in dB\footnote{The prediction $\hat{l}_{m,j}$ includes power calibration when needed. For example, the F-BBS model used in this work generates normalized channel hypotheses and therefore requires power calibration. The detailed procedure is provided in Section \ref{section5.B}.}, and $l_m = 10{\rm log}_{10} r_m$ be the reported RSRP measurement in dB. Thus, the prediction error can be expressed as
\begin{equation}\label{eq:error}
	\setlength\abovedisplayskip{3pt}%shrink space
	\setlength\belowdisplayskip{3pt}
	e_{m, j} = l_m - \hat{l}_{m, j}.
\end{equation}
It should be noted that both measurement noise and imperfections in the generated hypotheses can contribute to this error. Since the BS is assumed to have no prior knowledge of the measurement noise or operating SNR, we use a Gaussian surrogate to score the consistency of each prediction. Let $s_m > 0$ denote the effective prediction error scale at step $m$, common to all hypotheses. The setting of $s_m$ is detailed in Section \ref{section5.B}. The score for channel hypothesis $\hat{\bm{{\rm h}}}_j$ at step $m$ can thus be expressed as
\begin{equation}\label{eq:score}
	\setlength\abovedisplayskip{3pt}%shrink space
	\setlength\belowdisplayskip{3pt}
	L_{m, j} = {\rm e}^{-\frac{e_{m, j}^2}{2s_m^2}}.
\end{equation}
According to (\ref{eq:score}), a smaller absolute prediction error $\vert e_{m, j} \vert$ yields a higher score. The scale $s_m$ controls the tolerance for discrepancies between the predicted and measured RSRPs. Specifically, a smaller $s_m$ produces a stronger preference for hypotheses with smaller prediction errors, while a larger $s_m$ makes the weight update less sensitive to these differences. The score $L_{m, j}$ thus measures the consistency of hypothesis $j$ with the current RSRP observation, while the previous weight $\omega_j^{(m-1)}$ summarizes the evidence accumulated before step $m$. To combine the new observation with this accumulated evidence, the BS multiplies each previous weight by its corresponding score and normalizes the resulting weights as follows:
\begin{equation}\label{eq:update}
	\setlength\abovedisplayskip{3pt}%shrink space
	\setlength\belowdisplayskip{3pt}
	\omega_j^{(m)} = \frac{\omega_j^{(m-1)} L_{m, j}}{\sum_{l=1}^J \omega_l^{(m-1)} L_{m, l} }.
\end{equation}
If two channel hypotheses predict the same RSRP for the probed beam, they receive the same score in (\ref{eq:score}), and their relative weights remain unchanged. Otherwise, the new observation increases the relative weight of the hypothesis with the smaller prediction error. The BS also updates $\mathcal{I}_m = \mathcal{I}_{m-1} \cup \{ i_m \}$, and $a_j^{(m)} = \max\{a_j^{(m-1)}, G_{i_m, j}\}$. The updated weights and gains are then used in (\textbf{P4}) to select the next beam. After $M$ probes, the BS selects the probed beam with the highest reported RSRP for communication. The total probing overhead is still $Q+M$. The procedure of ACP is summarized in \textbf{Algorithm 2}. The procedures of GCP and ACP are illustrated in Fig. \ref{fig:gcp_and_acp}.

\begin{algorithm}[t]
	\caption{Adaptive Candidate Probing (ACP)}
	\label{alg:adaptive_probing}
	\begin{algorithmic}[1]
		\REQUIRE Wireless prompt $\bm{{\rm c}}$, initial RSRP observations
		$\mathbf o_Q$, trained generator $\mathcal G_{\vartheta}$,
		pool size $J$, and probing budget $M\leq J$.
		\ENSURE Probed candidate set $\mathcal B_M$ and communication
		beam $\mathbf w_{\mathrm{com}}$.
		
		\STATE Generate $\{\widehat{\mathbf h}_j\}_{j=1}^{J}$,
		recover $\mathcal C_J$, and compute $\mathbf G$
		as in steps 1--2 of \textbf{Algorithm~\ref{alg:greedy_selection}}.
		\STATE Initialize the RSRP predictor and its error scale
		using $\mathbf o_Q$.
		\STATE Set $\mathcal I_0=\emptyset$, $a_j^{(0)}=0$, and
		$\omega_j^{(0)}=1/J$ for all $j$.
		
		\FOR{$m=1,\ldots,M$}
		\FORALL{$i\in\{1,\ldots,J\}\setminus\mathcal I_{m-1}$}
		\STATE $\displaystyle
		d_i \gets \sum_{j=1}^{J}\omega_j^{(m-1)}
		\bigl(G_{i,j}-a_j^{(m-1)}\bigr)_+$
		\ENDFOR
		\STATE Choose $i_m\in
		\argmax_{i\notin\mathcal I_{m-1}}d_i$.
		\STATE Predict the RSRP $\widehat l_{m,j}$ for
		$\mathbf w_{i_m}$ under each hypothesis $j$,
		using the predictor state before probe $m$.
		\STATE Probe $\mathbf w_{i_m}$ and obtain the reported RSRP $l_m$.
		\FOR{$j=1,\ldots,J$}
		\STATE Obtain $e_{m,j}$ using (\ref{eq:error})
		\STATE Update the score $L_{m,j}$ using (\ref{eq:score})
		\ENDFOR
		\STATE Update $\omega_j^{(m)}$ using (\ref{eq:update}) for all $j$.
		\STATE $\mathcal I_m\gets\mathcal I_{m-1}\cup\{i_m\}$
		\STATE $a_j^{(m)}\gets
		\max\{a_j^{(m-1)},G_{i_m,j}\}$ for all $j$.
		\ENDFOR
		
		\STATE Form $\mathcal B_M=\{\mathbf w_{i_m}\}_{m=1}^{M}$.
		\STATE Choose $m^\star\in\argmax_{1\leq m\leq M}r_m$
		and set $\mathbf w_{\mathrm{com}}=\mathbf w_{i_{m^\star}}$.
	\end{algorithmic}
\end{algorithm}

\subsection{Complexity and Signaling-Latency Tradeoff}
\subsubsection{Computational Complexity}
As summarized in \textbf{Algorithm} \ref{alg:adaptive_probing}, ACP shares the same candidate generation, gain matrix calculation, and $M$-step greedy selection operations as GCP, incurring $\mathcal{O}(JC_{{\rm gen}} + J^2N_t + MJ^2)$ operations. Beyond GCP, ACP introduces two lightweight components: a one-time initialization of the RSRP predictor and error scale from $\bm{{\rm o}}_Q$, and feedback processing after each probe. Specifically, by caching $|\hat{\bm{{\rm h}}}_j^{\mathsf{H}}\bm{{\rm w}}_i|^2$ already evaluated during the construction of $\bm{{\rm G}}$, obtaining the RSRP predictions $\{\hat{l}_{m,j}\}_{j=1}^J$, computing the prediction errors in (\ref{eq:error}), evaluating the consistency scores in (\ref{eq:score}), and normalizing the hypothesis weights in (\ref{eq:update}) require only $\mathcal{O}(J)$ operations per probing step, totaling $\mathcal{O}(MJ)$ across all $M$ probes. Since $\mathcal{O}(MJ)$ is strictly dominated by the $\mathcal{O}(MJ^2)$, the adaptive updates do not increase the asymptotic complexity order. Consequently, ACP maintains the same asymptotic local computational complexity as GCP, i.e., $\mathcal{O}(JC_{{\rm gen}} + J^2N_t + MJ^2)$.

\subsubsection{Signaling Overhead and Latency Tradeoff}
While GCP and ACP incur the same total number of over-the-air probing transmissions $Q+M$, they differ fundamentally in their feedback structures, as illustrated in Fig. \ref{fig:gcp_and_acp}. Specifically, both i.i.d. probing and GCP perform non-adaptive candidate probing, i.e., the $M$ selected beams can be transmitted in a single burst (e.g., over consecutive CSI-RS occasions), requiring only one candidate feedback round from the UE to report the best beam index (or the $M$ RSRP values). In contrast, ACP operates in a closed-loop manner and requires $M$ sequential probe-and-report rounds, where the UE feeds back a scalar RSRP measurement after each probe to guide subsequent beam selection. Although each intermediate feedback message carries minimal payload, the sequential interactions introduce additional turnaround latency between consecutive probes. In section \ref{section5} we will show that GCP is preferable under dense initial probing or low SNR where feedback reweighting yields limited additional gain. Conversely, ACP is particularly useful when initial probing is sparse and RSRP feedback is reliable. Moreover, a hybrid batched-ACP strategy, i.e., probing a small batch of beams per feedback round, can naturally bridge GCP and ACP to balance adaptivity against turnaround latency, which is a promising solution.

\section{Numerical Results}\label{section5}
\subsection{Scenarios}
\begin{figure}[t]
	\centering
	\includegraphics[width=3.3in]{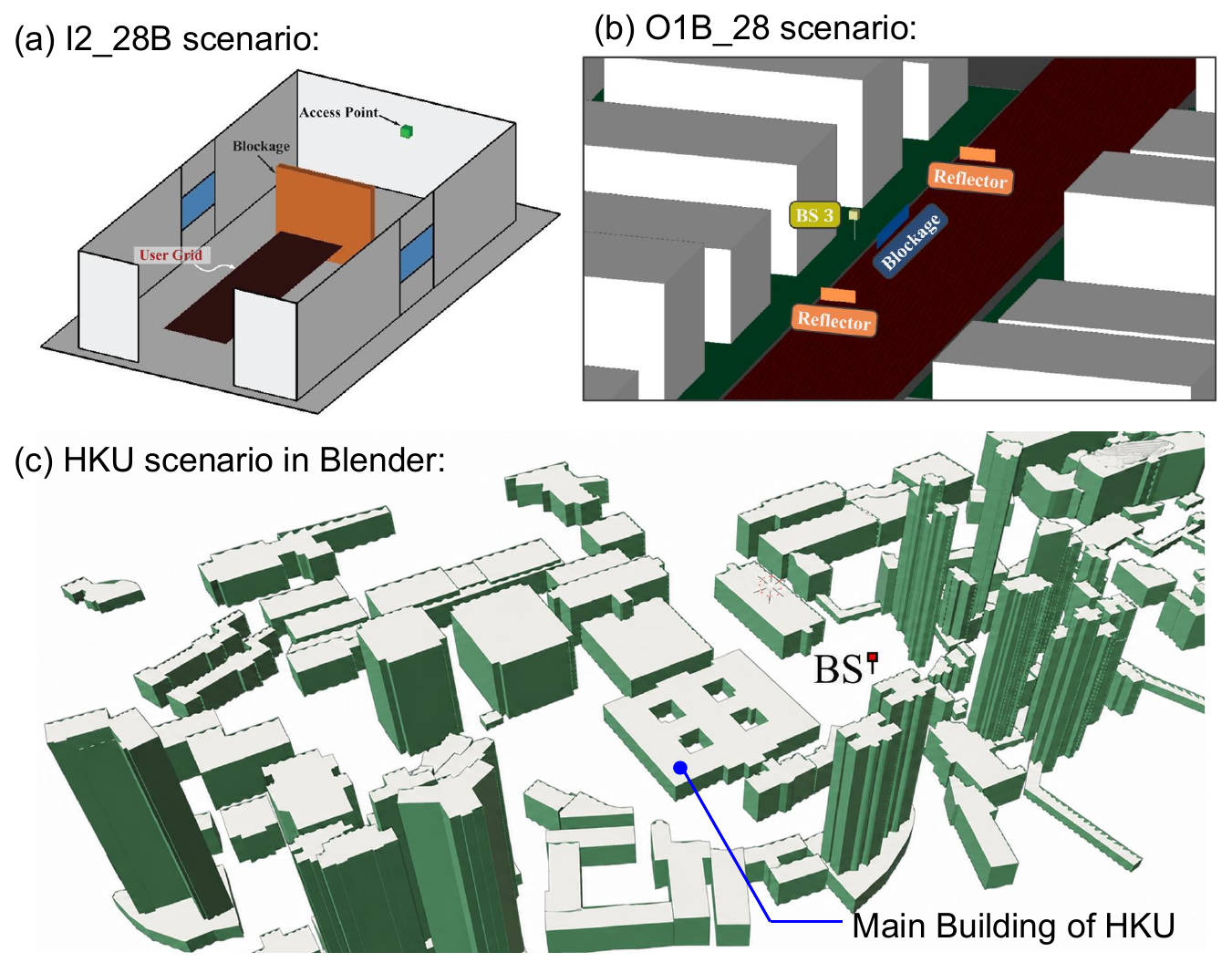}
	\caption{Scenarios used for evaluation: (a) indoor I2\_28B; (b) outdoor street O1B\_28; and (c) university campus HKU\_28.}
	\label{fig:scenarios}
\end{figure}

We evaluate GCP and ACP in the three 28-GHz RT scenarios used in F-BBS~\cite{zhou2026fastbeambrainstorm}. As shown in Fig.~\ref{fig:scenarios}, the scenarios include an indoor room, an outdoor street, and a university campus. The first two are public DeepMIMO scenarios~\cite{alkhateeb2019deepmimo}, and the third models the campus of The University of Hong Kong (HKU).

\begin{itemize}
	\item In I2\_28B, an indoor partition blocks the direct path from the wall-mounted BS to all 140,901 UE locations in the selected grid. This gives an entirely NLoS evaluation environment, as illustrated in Fig.~\ref{fig:scenarios}(a).
	
	\item O1B\_28 represents an outdoor street with obstructions and reflecting surfaces, as shown in Fig.~\ref{fig:scenarios}(b). We use BS~3 and user grid~1 of the DeepMIMO O1\_28 scenario. The selected grid contains 484,264 UE locations and includes both LoS and NLoS links.
	
	\item HKU\_28 covers a campus area with buildings of varying heights, streets, and open spaces. The scene in Fig.~\ref{fig:scenarios}(c) was built in Blender from OpenStreetMap data and ray-traced using Sionna RT. The BS is placed 15~m above ground, and the dataset contains 82,169 UE locations.
\end{itemize}

\subsection{Simulation Setup and Implementation Details}\label{section5.B}
The BS employs a 64-element ULA with half-wavelength antenna spacing. We adopt the F-BBS generator \cite{zhou2026fastbeambrainstorm}, with the model trained for each scenario using 80,000 channel samples. Each model is trained for 300 epochs with a batch size of 32 and a learning rate of $2 \times 10^{-4}$, and remains fixed throughout the evaluation. Performance is evaluated over 500 UEs in each scenario. The initial observations are obtained using $Q$ approximately uniformly spaced beams selected from a 64-beam DFT codebook. We compare GCP and ACP with independent sampling, which directly probes $M$ independently generated candidate beams. For each UE, all methods share the same initial observations and generated sample pool. The independent-sampling baseline uses the first $M$ samples, whereas GCP and ACP select $M$ candidates from the sample pool with $J (J>M)$ beams. After candidate probing, the beam with the highest reported RSRP is selected for communication. Performance is measured by its normalized beamforming gain, averaged over the testing UEs. 

We next specify the RSRP predictor and error scale used by ACP in Section \ref{section4.B}. Because F-BBS generates normalized channel hypotheses \cite{zhou2026fastbeambrainstorm}, their absolute power levels must be calibrated before their predicted RSRPs can be compared with the measured feedback. The initial probing observations provide a common reference for this calibration. For the generated channel hypothesis $\hat{\bm{{\rm h}}}_j$, define the uncalibrated power prediction in dB for beam $\bm{{\rm v}}_q$ as
\begin{equation}
	\setlength\abovedisplayskip{3pt}%shrink space
	\setlength\belowdisplayskip{3pt}
	u_{q,j} = 10{\rm log}_{10}\left(P_T \vert \hat{\bm{{\rm h}}}^{\mathsf{H}}_j \bm{{\rm v}}_q \vert^2\right), q=1,\cdots,Q.
\end{equation}
Let $l_q$ denote the RSRP measured in dB for $\bm{{\rm v}}_q$. Since weak responses are more susceptible to noise, we retain only those initial beams whose measured RSRPs are within 20 dB of the largest of the $Q$ measurements. Let $\mathcal{Q}_0$ denote the indices of the retained beams and $K = |\mathcal{Q}_0|$.  Therefore, we estimate the power offset for hypothesis $j$ as
\begin{equation}
	\setlength\abovedisplayskip{3pt}%shrink space
	\setlength\belowdisplayskip{3pt}
	b_j^{(0)} = \frac{1}{K} \sum_{q\in \mathcal{Q}_0 } (l_q - u_{q,j}), j=1,\cdots,J.
\end{equation}
After the initial calibration, the remaining error for beam $\bm{{\rm v}}_q$ can be expressed as $e_{q,j}^{(0)} = l_q - (u_{q,j} + b_j^{(0)}), q\in \mathcal{Q}_0$. These errors reflect differences in the predicted channel structure, as well as the measurement noise. In (\ref{eq:score}), we know that the error scale $s_m$ controls how strongly ACP penalizes an RSRP prediction at candidate probing step $m$. We first use the calibrated prediction to estimate a baseline error scale $s_{{\rm res}}$, which we then adjust according to the number of measurements available before candidate probing step $m$. To obtain $s_{{\rm res}}$, we compute the root mean square error residual over the $K$ retained beams for each hypothesis and take the median across the $J$ hypotheses, which can be expressed as
\begin{equation}
	\setlength\abovedisplayskip{3pt}%shrink space
	\setlength\belowdisplayskip{3pt}
	s_{{\rm res}} = \max\left\{s_{{\rm min}}, {\rm median}_{1\leq j \leq J} \sqrt{\frac{1}{K}\sum_{q\in \mathcal{Q}_0}(e^{(0)}_{q,j})^2}\right\}.
\end{equation}
The median reduces the influence of a few hypotheses with large errors. The lower bound $s_{{\rm min}}>0$ prevents an overly small scale from making the weight update too selective. In this work, we set $s_{{\rm min}}=6$ dB. We use $s_{{\rm res}}^2$ as an empirical estimate of the per-beam RSRP error variance. To set $s_m$ in (\ref{eq:score}), we consider a simple model in which, for a given channel hypothesis, the errors about its true power offset are independent across beams and have common variance $s_{{\rm res}}^2$. Before the $m$-th candidate probe, the BS estimates this offset by averaging the differences between measured and predicted RSRPs over $n=K+m-1$ previous beams. The offset estimate therefore has variance $s_{{\rm res}}^2 / n$. Since the error on the new probed beam is independent of the offset estimate, the prediction error has variance $s_{{\rm res}}^2 + s_{{\rm res}}^2 / n$. Thus, we have
\begin{equation}
	\setlength\abovedisplayskip{3pt}%shrink space
	\setlength\belowdisplayskip{3pt}
	s_m = s_{{\rm res}}\sqrt{1 + \frac{1}{K+m-1}}.
\end{equation}
Once the UE reports $l_m$, the BS computes the prediction error $e_{m,j} = l_m - \hat{l}_{m,j}$ as in (\ref{eq:error}) and updates the hypothesis weights using (\ref{eq:score}) and (\ref{eq:update}). It then incorporates the new measurement into each power offset through $b_j^{(m)} = b_j^{(m-1)} + \frac{e_{m,j}}{K+m}$. At the next step, ACP uses the updated weights to select the next candidate beam and the updated offsets to predict its RSRP.

\subsection{Illustrative Examples of GCP and ACP}

\begin{figure}[t]
	\centering
	\includegraphics[width=3.5in]{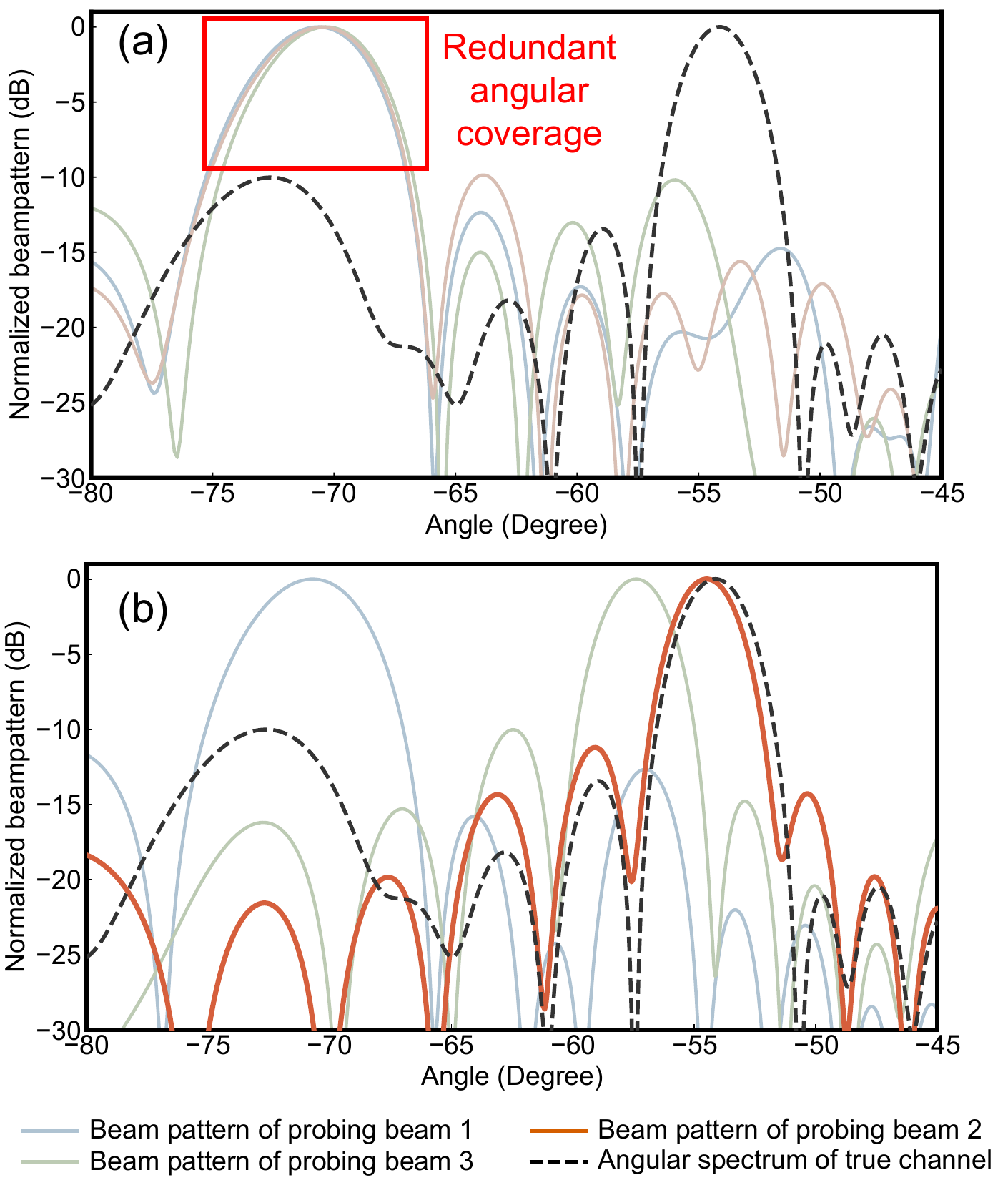}
	\caption{Beam patterns of three candidate beams selected by (a) i.i.d. probing and (b) GCP for a UE in I2\_28B, with $Q=9$ and $J=64$ under ideal probing.}
	\label{fig:intuitive_GCP_vs_iid}
\end{figure}

\begin{figure}[t]
	\centering
	\includegraphics[width=3.3in]{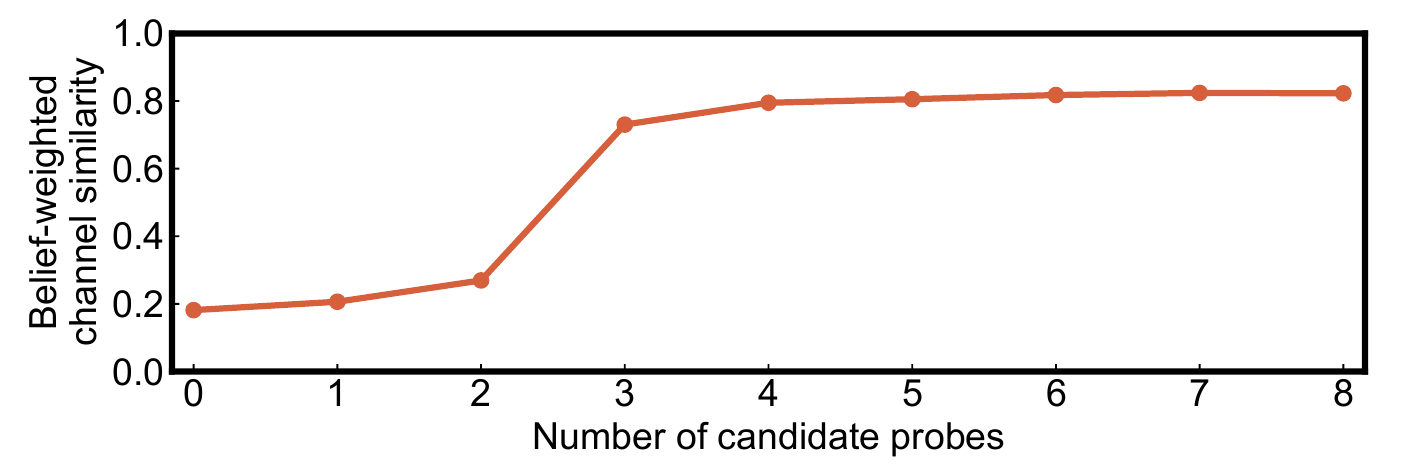}
	\caption{Evolution of the belief-weighted channel similarity during ACP for a UE in I2\_28B, with $Q=15$ and $J=64$ under ideal probing.}
	\label{fig:intuitive_ACP}
\end{figure}

In this subsection, we present two representative UE examples from I2\_28B to illustrate the underlying mechanisms of GCP and ACP. Fig. \ref{fig:intuitive_GCP_vs_iid} compares the beam patterns of $M=3$ candidate beams selected by i.i.d. probing and GCP for a UE with $Q=9$ and $J=64$. Under i.i.d probing in Fig.~\ref{fig:intuitive_GCP_vs_iid}(a), all three independently drawn beams concentrate their main lobes around $-70\degree$, causing redundant angular coverage while leaving the dominant channel direction near $-54\degree$ largely uncovered. By contrast, Fig.~\ref{fig:intuitive_GCP_vs_iid}(b) allocates the three probes to complementary angular directions, including one aligned with the dominant channel direction, increasing the achieved normalized gain from 0.102 to 0.873. This angular complementarity directly results from maximizing the marginal probing value, which penalizes gain overlap with previously selected beams and steers subsequent probes toward distinct plausible channel modes. Fig.~\ref{fig:intuitive_ACP} illustrates feedback-driven hypothesis reweighting for a UE with $Q=15$ and $J=64$. 
Specifically, we track the belief-weighted channel similarity after $m$ candidate probes, defined as $\sum_{j=1}^J \omega_j^{(m)} \frac{|\hat{\bm{{\rm h}}}_j^{\mathsf{H}} \bm{{\rm h}} |^2}{\Vert \hat{\bm{{\rm h}}}_j \Vert^2 \Vert \bm{{\rm h}} \Vert^2}$, which quantifies how closely the reweighted hypotheses align with the true channel $\bm{{\rm h}}$. Starting from 0.181 under the initial uniform weights, this similarity increases to 0.730 after 3 candidate probes. Since the generated candidate pool remains fixed, this substantial increase confirms that online RSRP feedback effectively downweights inconsistent hypotheses and concentrates belief on those matching the true channel, providing a refined basis for subsequent beam selection.

\subsection{Effect of the Candidate Probing Budget}
\begin{figure}[t]
	\centering
	\includegraphics[width=3.5in]{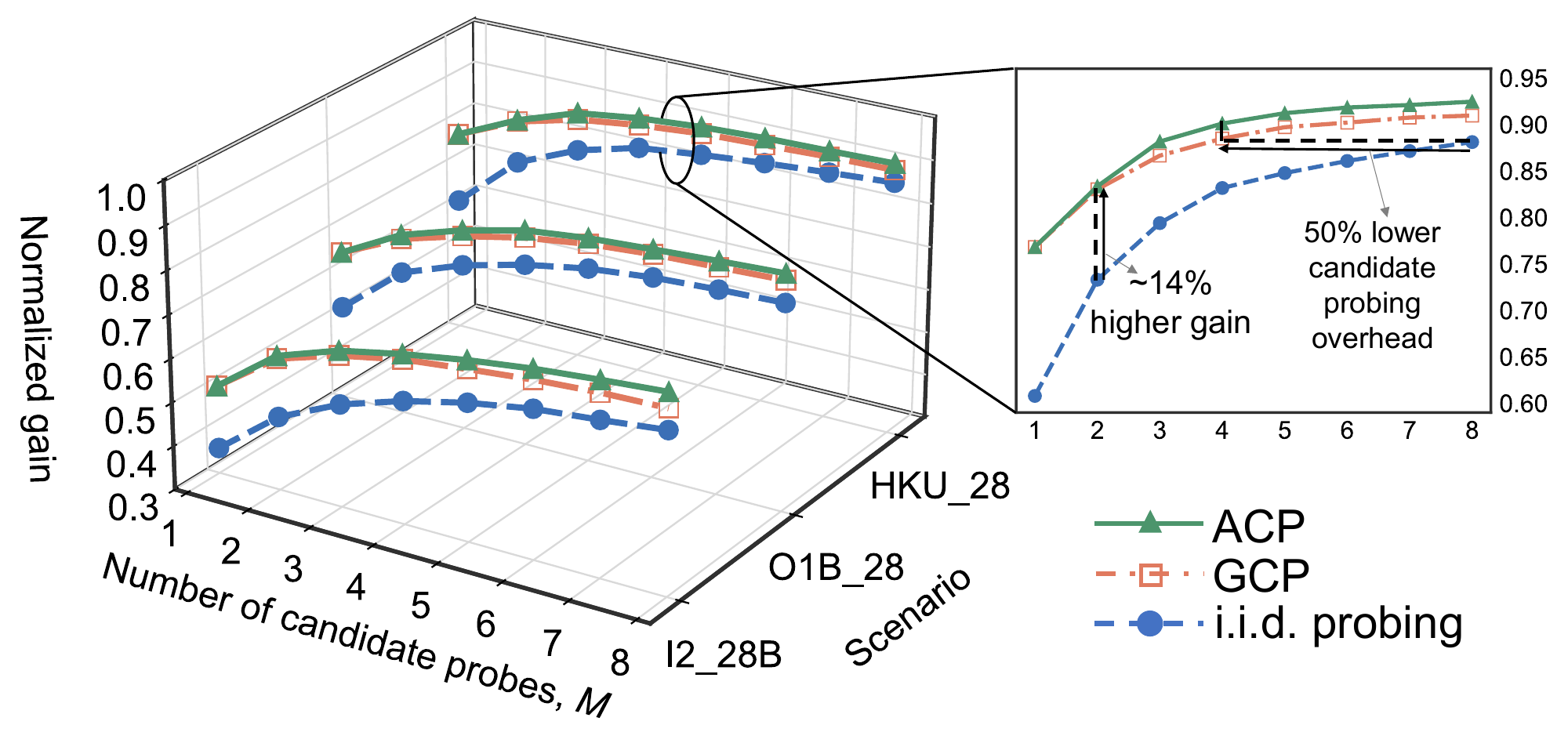}
	\caption{Mean normalized beamforming gain versus the candidate probing budget $M$ under noiseless probing, with $Q=15$ and $J=64$.}
	\label{fig:gain_vs_M_noiseless}
\end{figure}

Fig. \ref{fig:gain_vs_M_noiseless} shows the normalized beamforming gain performance as the number of candidate probes $M$ increases from 1 to 8, with $Q=15$ and $J=64$. Under noiseless probing, GCP and ACP achieve higher gains than i.i.d. sampling across all three scenarios. They select the same beam at $M=1$, since ACP starts with uniform hypothesis weights. Once candidate feedback becomes available, ACP can use it to guide subsequent probes. In HKU\_28, for example, ACP achieves a gain of 0.833 with two candidate probes, about 14\% higher than that achieved by i.i.d. sampling at the same budget. With 4 probes, ACP reaches 0.901, exceeding the 0.881 achieved by i.i.d. sampling with 8 probes while saving the candidate probing by 50\%.

\begin{figure*}[t]
	\centering
	\includegraphics[width=7.2in]{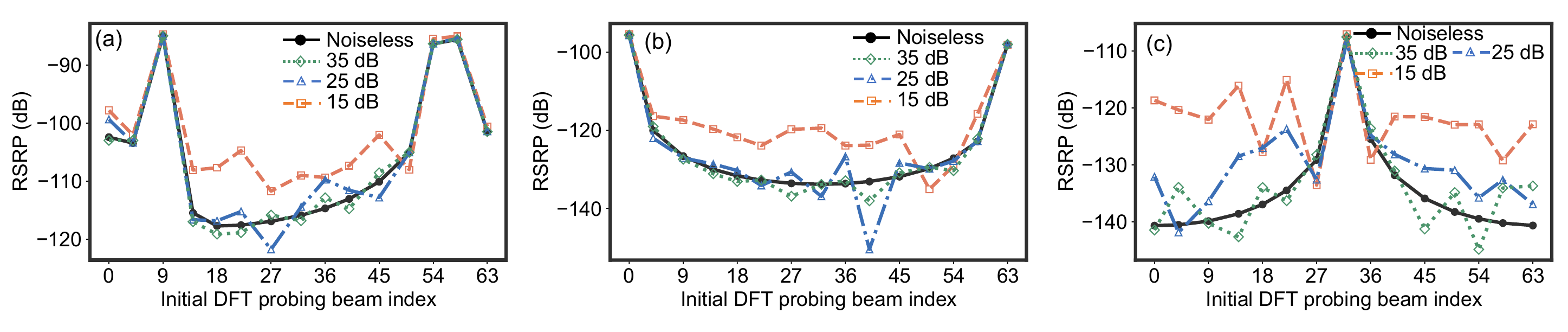}
	\caption{Initial DFT probing RSRP profiles under noiseless measurements and SNRs of 15, 25, and 35 dB: (a) I2\_28B; (b) O1B\_28; and (c) HKU\_28.}
	\label{fig:noisy_rsrp}
\end{figure*}

\begin{figure*}[t]
	\centering
	\includegraphics[width=7.2in]{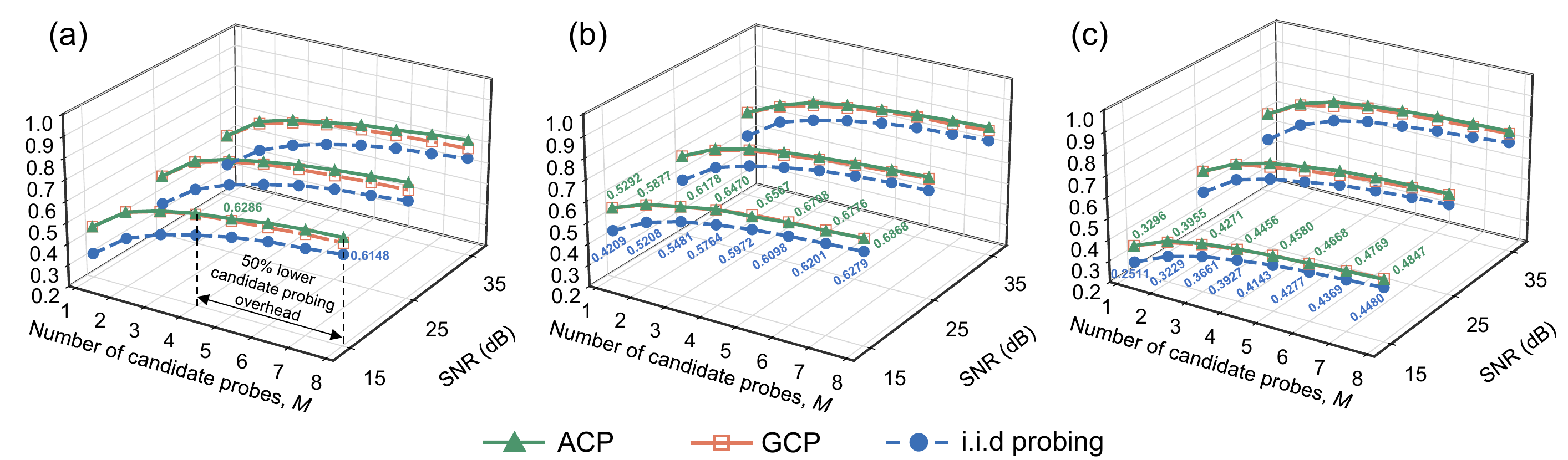}
	\caption{Mean normalized beamforming gain versus the candidate probing budget $M$ and SNR, with $Q=15$ and $J=64$: (a) I2\_28B; (b) O1B\_28; and (c) HKU\_28. Noise is applied to both initial and candidate probing stages.}
	\label{fig:gain_vs_M_noisy}
\end{figure*}

Next, we further evaluate the performance of each method under noisy probing conditions, including both initial DFT probing and the subsequent candidate probing. For each scenario, we define the reference power $P_{{\rm ref}}$ as the mean noiseless received power over the tested UEs and all 64 beams in the DFT codebook. For a nominal SNR of $\gamma$ dB, we set the noise variance to $\sigma^2 = P_{{\rm ref}} 10^{-\gamma / 10}$ and use this variance for both initial and candidate probing. For each probe, we add circularly symmetric complex Gaussian noise $n\sim \mathcal{CN}(0, \sigma^2)$ to the noiseless complex response $s$ before computing the reported RSRP as $|s + n|^2$. The communication beam is then selected using this reported noisy RSRP. Fig. \ref{fig:noisy_rsrp} first illustrates the impact of different SNRs on the RSRP. As the noise level increases, the RSRP exhibits a larger deviation, making erroneous decisions more likely. Fig. \ref{fig:gain_vs_M_noisy} demonstrates the resulting normalized beamforming gains when the same nominal SNR is applied to both initial and candidate probing. It can be observed that the advantages of GCP and ACP over i.i.d. sampling remain significant under noisy probing. For instance, in I2\_28B at 15 dB, ACP reaches a gain of 0.629 with 4 probes, compared with 0.615 for i.i.d. sampling with 8 probes.

\subsection{Effect of the DFT Probing Budget}
\begin{figure*}[t]
	\centering
	\includegraphics[width=7.2in]{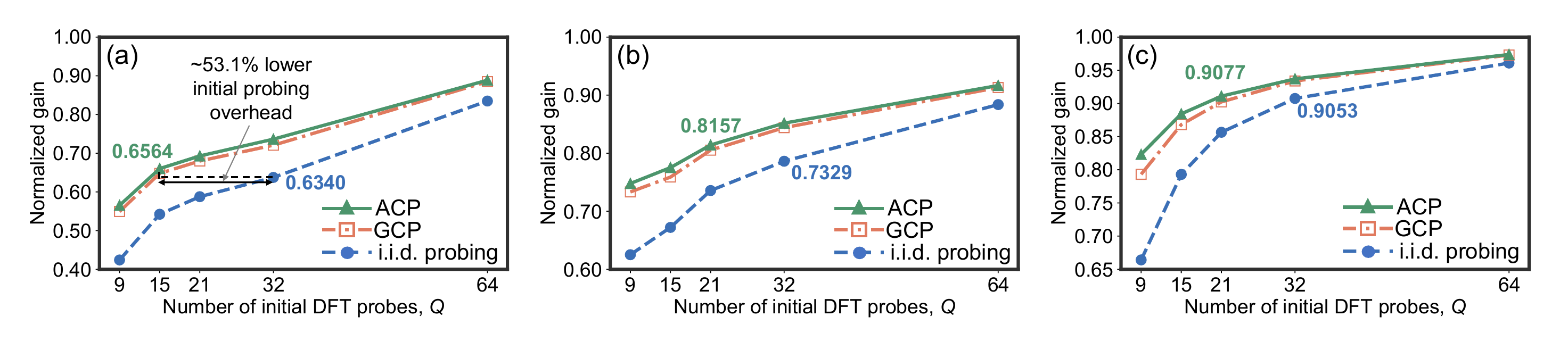}
	\caption{Mean normalized beamforming gain versus the initial DFT probing budget $Q$ under noiseless probing, with $M=3$ and $J=64$: (a) I2\_28B; (b) O1B\_28; and (c) HKU\_28.}
	\label{fig:gain_vs_Q_noiseless}
\end{figure*}

\begin{figure*}[t]
	\centering
	\includegraphics[width=7.2in]{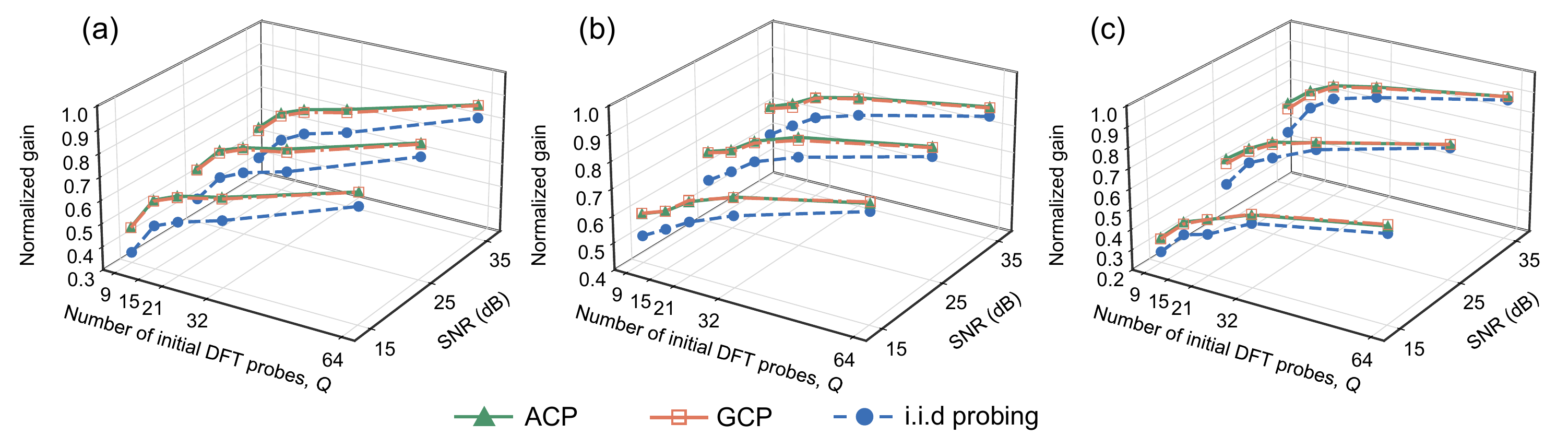}
	\caption{Mean normalized beamforming gain versus the initial DFT probing budget $Q$ and SNR, with $M=3$ and $J=64$: (a) I2\_28B; (b) O1B\_28; and (c) HKU\_28. Noise is applied to both initial and candidate probing stages.}
	\label{fig:gain_vs_Q_noisy}
\end{figure*}

In GenSSBF, we know that increasing the number of initial probes $Q$ gives the generator more observations to condition on, which can improve beamforming performance. We examine whether the proposed GCP and ACP can achieve a given beamforming gain with fewer initial probes than i.i.d. sampling. To this end, we vary $Q$ while fixing the candidate probing budget at $M=3$ and the pool size at $J=64$. Fig. \ref{fig:gain_vs_Q_noiseless} shows the results under the noiseless setting. It can be observed that GCP and ACP achieve higher normalized gains than i.i.d. sampling at the same $Q$, with the gains over i.i.d. sampling generally increasing as $Q$ decreases. This advantage leads to fewer initial probes being required to achieve a given gain. For example, in I2\_28B, GCP and ACP achieve gains of 0.645 and 0.656, respectively, with $Q=15$. Both exceed the gain of 0.634 achieved by i.i.d. sampling with $Q=32$, using 53\% fewer initial probes at the same candidate probing budget. Fig. \ref{fig:gain_vs_Q_noisy} shows that GCP and ACP can reduce the initial probing budget even when both probing stages are noisy. In I2\_28B at 15 dB, GCP and ACP achieve mean normalized gains of 0.595 and 0.599, respectively, with 15 initial probes. Both exceed the gain of 0.532 achieved by i.i.d. sampling with 21 initial probes, saving 6 initial measurements. It should be noted that the additional gain of ACP over GCP depends on the \textit{residual uncertainty left by the initial observations} and \textit{the amount of useful information provided by the subsequent feedback}. For instance, in HKU\_28, at an SNR of 35 dB, the gain advantage of ACP over GCP falls from 0.030 at $Q=9$ to less than 0.001 at $Q=64$. This suggests that a denser initial prompt leaves less room for feedback reweighting to improve beam selection. With $Q$ fixed at 9, the advantage of ACP over GCP grows from 0.006 to 0.030 as the SNR increases from 15 to 35 dB, which means that reliable feedback is important for ACP.

\subsection{Effect of the Candidate Pool Size}
The preceding simulations varied the number of over-the-air probes. Here, we study how the size of the locally generated candidate pool affects beamforming performance at a fixed probing budget. We vary $J$ from 16 to 64 with $Q=15$ and $M=3$. A larger pool gives GCP and ACP more beams to choose from when allocating the three candidate probes. The i.i.d. baseline probes three draws directly and yields essentially the same gain across the tested pool sizes.

Table~\ref{tab:gain_vs_pool_size} shows that increasing $J$ from 16 to 64 improves the mean normalized gain of both GCP and ACP in all three scenarios under both noiseless and noisy probing. But it should be noted that a larger pool also takes longer to generate. The mean local computation time of ACP across the three scenarios is 11.0, 17.4, and 33.3~ms per UE for $J=16$, $32$, and $64$, respectively, on an NVIDIA GeForce RTX 4060 Laptop GPU and an Intel Core i7-14650HX CPU. The benefit of this extra computation varies by scenario. At an SNR of 25 dB, increasing $J$ from 32 to 64 raises the ACP gain from 0.630 to 0.647 in I2\_28B. At the same SNR level, in HKU\_28, $J=32$ already achieves a gain of 0.604, close to the 0.606 obtained with $J=64$, while reducing local computation time from 33.8 to 17.7~ms per UE.

\begin{table*}[t]
	\centering
	\caption{Mean normalized beamforming gain for different candidate pool
		sizes $J$ with $Q=15$ and $M=3$. Noise is applied to both initial and candidate probing stages.}
	\label{tab:gain_vs_pool_size}
	\small
	\setlength{\tabcolsep}{4pt}
	\begin{tabular}{@{}ll *{3}{rrrr}@{}}
		\toprule
		& & \multicolumn{4}{c}{$J=16$}
		& \multicolumn{4}{c}{$J=32$}
		& \multicolumn{4}{c}{$J=64$} \\
		\cmidrule(lr){3-6}
		\cmidrule(lr){7-10}
		\cmidrule(l){11-14}
		Scenario & Method
		& Noiseless & 35 dB & 25 dB & 15 dB
		& Noiseless & 35 dB & 25 dB & 15 dB
		& Noiseless & 35 dB & 25 dB & 15 dB \\
		\midrule
		I2\_28B & GCP
		& 0.616 & 0.608 & 0.598 & 0.548
		& 0.631 & 0.623 & 0.624 & 0.574
		& 0.645 & 0.641 & 0.635 & 0.595 \\
		& ACP
		& 0.613 & 0.609 & 0.595 & 0.555
		& 0.644 & 0.637 & 0.630 & 0.580
		& 0.656 & 0.654 & 0.647 & 0.599 \\
		\addlinespace[2pt]
		O1B\_28 & GCP
		& 0.730 & 0.707 & 0.667 & 0.593
		& 0.750 & 0.727 & 0.681 & 0.607
		& 0.750 & 0.726 & 0.691 & 0.616 \\
		& ACP
		& 0.736 & 0.707 & 0.664 & 0.594
		& 0.757 & 0.727 & 0.691 & 0.600
		& 0.763 & 0.740 & 0.699 & 0.615 \\
		\addlinespace[2pt]
		HKU\_28 & GCP
		& 0.854 & 0.687 & 0.585 & 0.401
		& 0.863 & 0.697 & 0.592 & 0.418
		& 0.866 & 0.699 & 0.592 & 0.424 \\
		& ACP
		& 0.865 & 0.694 & 0.593 & 0.401
		& 0.874 & 0.705 & 0.604 & 0.416
		& 0.882 & 0.720 & 0.606 & 0.434 \\
		\bottomrule
	\end{tabular}
\end{table*}

\section{Conclusion}\label{section6}
This paper investigated strategic candidate probing for GenSSBF under a limited probing budget. We first reformulated beam brainstorming as a candidate selection problem and derived a marginal probing value that quantifies each candidate's additional gain beyond those already selected. This criterion led to a plug-and-play GCP algorithm, and an ACP algorithm that incorporates probing feedback into subsequent candidate selection. Simulations across three scenarios under noiseless and noisy probing demonstrated improved beamforming gain over independent sampling at the same probing budget. The results highlight that effective beam brainstorming requires candidates that provide complementary gains across plausible channels. They further show that feedback adaptation is particularly beneficial when initial sensing leaves substantial channel uncertainty and subsequent measurements are reliable. Overall, the proposed GCP and ACP offer promising ways to make more effective use of the limited probing resource in GenSSBF.

\newpage

\vspace{11pt}

\vfill

\end{document}